\documentclass[aps,twocolumn,graphics,floatfix,tightenlines]{revtex4-2}
\usepackage{graphics}
\usepackage{epstopdf}
\usepackage{epsfig}
\usepackage{graphicx}
\usepackage{epsf,epic}
\usepackage{color}
\usepackage{subfig}
\usepackage{mathtools}
\usepackage{amsmath}
\usepackage{booktabs}
\usepackage{multirow}
\usepackage{physics}
\usepackage{hyperref}
\usepackage{amsfonts}
\usepackage{wrapfig}
\usepackage{breqn}
\usepackage{pstricks}
\usepackage[utf8x]{inputenc}
\usepackage{multirow}
\usepackage{fancyref}
\usepackage{pst-node}
\usepackage{soul}
\usepackage{float}
\usepackage{bm}
\usepackage{dcolumn}
\newcommand{\etal}{\textit{et al.\ }}
\newcommand{\ie}{\textit{i.e.\ }}

\makeatletter
\usepackage{etoolbox} 
\appto{\appendix}{%
	\@ifstar{\def\theequation@prefix{A.}}%
	{}%
}
\preto\maketitle{%
	\begingroup\lccode`~=`,
	\lowercase{\endgroup
		\let\saved@breqn@active@comma~
		\let~}\active@comma 
}
\appto\maketitle{%
	\begingroup\lccode`~=`,
	\lowercase{\endgroup
		\let~}\saved@breqn@active@comma 
}
\makeatother
\begin{document}
	
	\title{Electronic structure and exchange interactions in altermagnetic MnGeP$_2$ in the quasiparticle-self-consistent $GW$ approach}\
	\author{Ilteris K. Turan}
	\email{ilteris.turan@case.edu}
	\author{Walter R L. Lambrecht}
	\email{walter.lambrecht@case.edu}
	\affiliation{Department of Physics, Case Western Reserve University, 10900 Euclid Avenue, Cleveland, OH 44106-7079, USA}
	\author{Jerome Jackson}
	\email{jerome.jackson@stfc.ac.uk}
	\affiliation{Scientific Computing Department, STCF Daresbury Laboratory, Warrington WA4 4AD, United Kingdom}
	
	\begin{abstract}
	  The quasiparticle self-consistent (QS) $GW$ method is used to study the electronic band structure, optical dielectric function, and exchange interactions in chalcopyrite, $I\bar{4}2d$, structure MnGeP$_2$. The material is found to be an antiferromagnetic semiconductor with lowest direct gap of 2.44 eV at the $\Gamma$ point and a lower indirect gap of 1.87 eV from $\Gamma$ to $M$. The material is an altermagnet because the two magnetic atoms of opposite spin are related by a two-fold rotation operation perpendicular to the main 4-fold rotation inversion axis. The spin splittings along a low symmetry line like $PN$ is sizable while at {\bf k}-points on the diagonal mirror planes or on the twofold symmetry axes the spin splitting is zero. The exchange interactions are calculated using a linear response approach. The antiferromagnetic exchange-interaction between nearest neighbors in the primitive unit cell is dominating and found to be slightly decreasing upon carrier doping but not sufficiently to change the interaction to become ferromagnetic. The bare (non-interacting) and interacting transverse spin susceptibilities, which provide interatomic site exchange interactions after averaging over the muffin-tin spheres, are calculated from the $GW$ band structure and wave functions. From these exchange interactions, the spin wave spectra are obtained along the high symmetry lines and  the N\'{e}el temperature is calculated using the mean-field, and Tyablikov estimations. The dielectric function and the optical absorption spectra are calculated including excitonic effects using the Bethe Salpeter equation. The exchange interactions around Mn$_{\rm Ge}$ defect sites is also studied. While we find it can generate ferromagnetic interactions with neighboring spins, we did not find direct evidence of producing an overall ferromagnetic phase. First, if Mn antisites are introduced by exchanging Mn with a nearby Ge, the interactions stay largely antiferromagnetic. Second, when we add additional Mn, in other words in Mn-rich stoichiometry, the Mn$_{\rm Ge}$ antisites produce a strong ferromagnetic interaction primarily with the Mn in the same basal plane but weaker ferromagnetic interaction with adjacent plane Mn. The interactions between regular lattice Mn stay antiferromagnetic as before and thus favor keeping the antiferromagnetic order along the [001] direction. Adding Mn antisites however does lead to a metallic band structure.

	\end{abstract}
	\maketitle
	
	\section{Introduction}
	Room-temperature magnetic semiconductors would greatly enhance the field of spintronics by the possibility of controlling both the magnetic moment and the carrier concentration
	but have remained elusive so far\cite{Sato2010}. One possible route proposed in the past is to dope chalcopyrite semiconductors like ZnGeP$_2$ or CdZneP$_2$ with a magnetic transition metal element like Mn \cite{Medvedkin2000,Medvedkin2002}. Compared to doping Mn in GaAs or other III-V compounds, this has the advantage that Mn is a divalent element and can therefore be more readily introduced in a II-IV-V$_2$ compound. Doping of CdGeP$_2$ and ZnGeP$_2$ were reported to show ferromagnetic behavior \cite{Medvedkin2000,Sato2001,Medvedkin2002,Sato2003,Baranov2003,Popov2004,Ishida2003,Medvedkin2024,Medvedkin2024b} but density functional calculations predicted antiferromagnetic order instead \cite{ZhaoYJ2001,Mahadevan2003}.
	Pure MnGeP$_2$ and MnGeAs$_2$ were also synthesized and grown by various
	epitaxial growth methods\cite{Sato2005,Cho2004} and found to exhibit ferromagnetic hysteresis. Recently, ferromagnetic resonance was observed in MnGeP$_2$ bulk single crystals grown by a gradient freeze method \cite{Bardeleben2024}.
	However, the origin of the ferromagnetism in these systems is not clear.
	Density functional calculations found these systems to be antiferromagnets and
	Mahadevan and Zunger \cite{Mahadevan2003} explained that Mn in II-VI
	(and similarly in II-IV-V$_2$) compounds are expected to prefer antiferromagnetism because in the ferromagnetic state, there are no empty Mn d-orbitals on a neighboring Mn to hop onto from a given Mn atom. Based on their defect studies, they proposed that Mn$_{\rm Ge}$ anti-site defects could be responsible for the ferromagnetism. More generally, one might wonder whether carrier-mediated exchange interactions could tilt the balance in favor of ferromagnetism. 
	
	All previous computational studies of these materials were carried out within the framework of density functional theory (DFT), which typically underestimates band gaps and may underestimate magnetic moments. 
	Here we present a study based on the quasiparticle-self-consistent $GW$  (QS$GW$) method,\cite{MvSQSGWprl,Kotani07}  which is based on the many-body perturbation theoretical (MBPT) approach of Hedin\cite{Hedin65,Hedin69} which evaluates the dynamical self-energy $\Sigma=iGW$ in terms of the one-electron Green's function $G$ and the screened Coulomb interaction $W$. Besides accurately predicting the band gaps, this method can also be extended to a linear response approach to extract the exchange interactions from the transverse spin susceptibility $\chi^{+-}$ \cite{Kotani2008}. This method is used here to predict the Heisenberg-type exchange interactions without  making {\sl a-priori} assumptions on the type of ordering. Starting from a spin-polarized ferromagnetic reference
        state,  the thus calculated 
      exchange interactions already indicate a strong antiferromagnetic coupling between the two Mn atoms in the primitive unit cell instead of the starting ferromagnetic alignment.  This is the dominant exchange interaction found. Other longer range interactions between cells are much smaller. Based on this prediction,
      we then calculate the antiferromagnetic state within the same unit cell. The exchange interactions still predict antiferromagnetic coupling but are now more reliably calculated. We calculate the N\'{e}el temperature in the mean field approximation and using the Tyablikov method \cite{Tyablikov1959,Callen63,Rusz2005} and study the spinwaves. 
      To complete our study, the optical dielectric function is calculated in the AFM and FM states using the Bethe-Salpeter equation (BSE) approach.
      Interestingly, the band structure calculations reveal that MnGeP$_2$ is an altermagnet with sizable spin splittings along low symmetry lines in the Brillouin zone. Symmetry analysis confirms that the $I\bar{4}2d$ structure indeed is consistent with altermagnetism. While our calculations of pure MnGeP$_2$ thus confirm previous findings from DFT, it reveals a new interesting aspect and predicts the band gaps more accurately. In the remainder of the paper we investigate possible defect related origins for the observation of ferromagnetism.

To study the possibility of carrier-mediated ferromagnetism we essentially use a rigid band approach in which hole doping is introduced by changing the valence band filling compensated by a neutralizing homogeneous background. Similarly, we can also add n-type doping in the conduction band. Either of these methods are found to reduce the nearest neighbor antiferromagnetic exchange interaction but the effect is too small to flip the balance toward net ferromagnetism even for an unrealistically high doping concentration. On the other hand, the exchange interactions between Mn$_{\rm Ge}$ antisites and neighboring regular Mn atoms are found to be ferromagnetic and thus more promising as a  possible route toward ferromagnetism. We studied these antisites in 16 atom and 64 atom supercells. We note, however, that in the former case this corresponds to a layer of MnP. Precipitates of MnP have also been invoked as a source of ferromagnetism in this system \cite{Bardeleben2024} but would differ from the present model by adopting a different local crystal structure.  We find that while Mn$_{\rm Ge}$
antisites cause some local ferromagnetic interactions to their neighbors if there is excess Mn, but not if we simply swap Mn with a neighboring Ge, the system maintains overall its antiferromagnetic ordering along [001] planes. Thus, this path to ferromagnetism based on point defects is also ruled out and the only
experimentally confirmed origin of ferromagnetism in actual samples appears to be the presence of a MnP secondary phase.

	\begin{figure}
		\includegraphics[width=8.5cm]{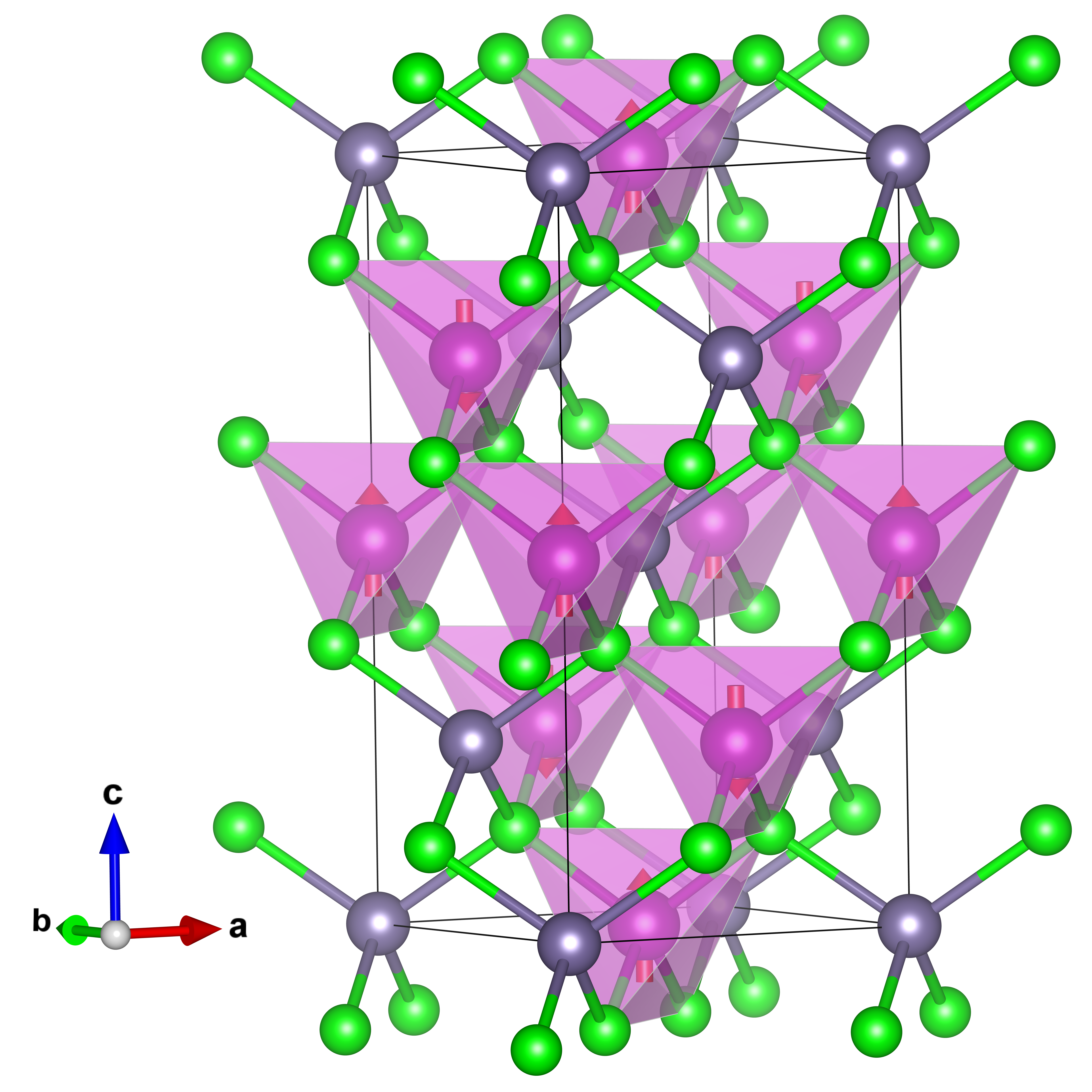}
		\caption{Crystal structure of the conventional cell with the AFM spin arrangements on the Mn sites in pink, along with Ge sites in gray and P sites in green. Pink areas are the nearest neighbor tetrahedra surrounding each Mn. The image is generated using the VESTA 3 software \cite{Vesta}.
			\label{struc}}
	\end{figure}
	
	\section{Computational Methods}\label{sec:method}
	\subsection{Quasiparticle self-consistent $GW$ method}\label{sec:gw}
	The calculations are performed using the {\sc Questaal} suite of codes which implements both DFT and MBPT using a full-potential linear muffin-tin orbital (FP-LMTO) basis set \cite{questaalpaper}.
	The QS$GW$ method is a variant of the $GW$ approach in which an improved
	non-local (but energy independent) exchange correlation potential $\tilde\Sigma_{xc}$ is extracted from the dynamic (\ie energy-dependent) self-energy,
	$\Sigma(\omega)$ as follows
	\begin{equation}
		\tilde{\Sigma}_{xc}=\frac{1}{2}{\rm Re}\left\{\sum_{nm}|\psi_n\rangle[\Sigma_{nm}(\epsilon_n)+\Sigma_{nm}(\epsilon_m)]\langle\psi_m|\right\},\label{eqvxc}
	\end{equation}
	in which ${\rm Re}$ stands for Hermitian part \cite{Kotani07} and $\epsilon_n$, $|\psi_n\rangle$ are the eigenvalues and eigenstates of the $H^0$ Hamiltonian. The $H^0$ is an effective one-particle Hamiltonian which initially is at the DFT level  but which is subsequently adjusted in the quasiparticle self-consistency loop.  
	For details of the justification of this approach see \cite{Kotani07,Ismail-Beigi_2017}. Thus, instead of making the Dyson equation Green's function $G=G^0+G^0\Sigma G$ self-consistent, it adjusts the $G^0$  corresponding to the  effective Hamiltonian $H^0$ so as to make the perturbation  by the interactions that go beyond an effective potential and are treated by MBPT as small as possible. The procedure is as follows: from a given $H^0$'s eigenstates and eigenvalues, the one particle Green's function $G^0$ and the polarization propagator  $P^0=-iG^0G^0$ are calculated and provide the screened Coulomb interaction $W^0=(1-P^0v)^{-1}v$ and eventually the self-energy   $\Sigma(\omega)=iG^0W^0$ is calculated as a matrix in the basis of $H^0$ eigenstates. Then $\tilde{\Sigma}_{xc}$ is extracted from Eq.(\ref{eqvxc}) and its difference from the original DFT $v_{xc}$ is added to $H^0$ for the next iteration. This procedure is iterated until the change in  $\tilde\Sigma$ from one iteration to the next falls below a tolerance. At self-consistency,  the band structure becomes independent of the DFT starting $H^0$ and the quasiparticle energies equal the Kohn-Sham eigenvalues of $H^0$. Thus, the method focuses on obtaining the real part of the quasiparticle excitation energies or band structure, rather than the full dynamic self-energy or quasiparticle lifetimes.

        The optical dielectric function, which corresponds to the macroscopic ${\bf q}\rightarrow0$ limit of the dielectric response is calculated  using the BSE, which incorporates both local-field and electron-hole interaction effects.\cite{Strinati1988,Onida02}  This method was recently implemented in the {\sc Questaal} suite of codes using the same product basis set approach as the $GW$ implementation \cite{Cunningham18}. The BSE  can also be applied to the ${\bf q}$-dependent $W({\bf q},\omega)$ as a way to go beyond the random phase approximation (RPA)
	in the screening of $W$ by incorporating electron-hole effects \cite{Cunningham18,Cunningham23}. This approach is equivalent to a vertex correction and amounts to incorporating ladder-diagrams in $W({\bf q},\omega)$ and overcomes the overestimate of band gaps by QS$GW$ due to under-screening that has become apparent from numerous applications of the QS$GW$ method. Indeed it has become clear that the original QS$GW$ systematically overestimates band gaps and underestimates dielectric constants.\cite{Bhandari18}  
        This method was dubbed the QS$G\hat W$ method in \cite{Cunningham23}
        but we will here refer to it as QS$GW^{BSE}$ as opposed to the original QS$GW^{RPA}$. 

        Technical details of the QS$GW$ implementation can be found in \cite{Kotani07}. Its hallmark is the use of a mixed-product basis set to represent two-particle quantities, an approach which was first introduced by Aryasetiawan and Gunnarsson \cite{Aryasetiawan94}. Its advantage lies in the use of products of highly accurate numerical radial function times spherical harmonic basis functions or partial waves inside the muffin-tin spheres to represent the short-range screening response. This is easier to converge than a plane-wave representation, which would require very high wave vectors. In the sum over states in the evaluation of the self-energy we take into account all the bands spanned by the muffin-tin orbital basis set which extend up to about 6 Ry above the Fermi energy or
        190 empty bands and provides adequate convergence. 
	The basis set used here was tested for convergence and consists of $spdfspd$ functions including Ge-$3d$ local orbitals. The self-energy matrix is approximated by a diagonal average value above a cut-off of 2.0 Ry. A well converged $5\times5\times5$ {\bf k}-mesh is used to sample the Brillouin zone both for charge self-consistent calculations and to sample the self-energy. 
	
	\subsection{Magnetic Susceptibility and Heisenberg Exchange Interactions} \label{sec:chi}
To determine the magnetic ground state, the critical temperature and the low energy excitations (spin waves), one usually starts from a Heisenberg-type spin-Hamiltonian, with the exchange interaction parameters determined from first-principles calculations. The latter can be obtained  in various ways. The first of those is to map the total energy differences of various magnetic configurations calculated within spin-dependent DFT to a Heisenberg type spin Hamiltonian with a predetermined extent of the interactions and number of independent exchange interaction parameters. The second approach is to use linear response around a chosen reference spin configuration. This approach has mostly been used in the context of the multiple scattering or Green's function description of the underlying spin-dependent electronic structure and estimates the first-order corrections to the total energy to a static infinitesimal rotation of the spins within each muffin-tin sphere \cite{Liechtenstein87}. The generalization of this approach in MBPT is to calculate the dynamic transverse spin susceptibility \cite{Aryasetiawan99}. The spin wave (SW) energies are then in principle obtained from the poles of this transverse spin susceptibility, which one can either calculate in the bare or the interacting limit and Heisenberg type exchange interactions can still be obtained from a static limit. The dynamic spin susceptibility furthermore contains information on the lifetime of the SW excitations.  

Here, we use the latter approach in the framework of the QS$GW$ method. This linear response approach starts from the transverse spin susceptibility, $\chi_{\bf q}^{0+-}({\bf r},{\bf r}',\omega)$, expressed in terms of the QS$GW$ wave functions and energy levels,
\begin{equation}
		\begin{multlined}
			\chi_{\bm{q}}^{0+-}(\bm{r}, \bm{r}',\omega)=\\ \sum_{\bm{k}n\downarrow}^{occ.}\sum_{\bm{k}'n'\uparrow}^{unocc.} \frac{\psi^{*}_{\bm{k}n\downarrow}(\bm{r}) \psi_{\bm{k}'n'\uparrow}(\bm{r})\psi^{*}_{\bm{k}'n'\uparrow}(\bm{r}')\psi_{\bm{k}n\downarrow}(\bm{r}')}{\omega-(\epsilon_{\bm{k}'n'\uparrow}-\epsilon_{\bm{k}n\downarrow})+i\delta}\\
			+\sum_{\bm{k}n\downarrow}^{unocc.}\sum_{\bm{k}'n'\uparrow}^{occ.}\frac{\psi^{*}_{\bm{k}n\downarrow}(\bm{r}) \psi_{\bm{k}'n'\uparrow}(\bm{r})\psi^{*}_{\bm{k}'n'\uparrow}(\bm{r}')\psi_{\bm{k}n\downarrow}(\bm{r}')}{-\omega-(\epsilon_{\bm{k}n\downarrow}-\epsilon_{\bm{k}'n'\uparrow})+i\delta},\label{eq:chi}
		\end{multlined}	
	\end{equation} 
where $\bm{k}'=\bm{k}+\bm{q}$, and $\psi_{n{\bf k}\sigma}$ and
$\epsilon_{n{\bf k}\sigma}$ are the spin-dependent band structure eigenstates and eigenvalues, which in our case are calculated in the QS$GW^{BSE}$ approximation.
	
        From $\chi_{\bf q}^{0+-}({\bf r},{\bf r}',\omega)$, a coarse-grained version between magnetic sites, $D^{0+-}_{ij}({\bf q},\omega)$, is obtained by a projection onto the spheres within a rigid spin approximation, as follows \cite{Kotani2008}, 
        \begin{equation}
D^{0+-}_{ij}({\bf q},\omega)=\int_{s_i}d^3r\int_{s_j}d^3r'\bar{e}_i({\bf r})\chi_{\bf q}^{0+-}({\bf r},{\bf r}',\omega))\bar{e}_j({\bf r}'),
\end{equation}
  with $e_i({\bf r})=M_i({\bf r})/M_i$, $M_i=\int_{s_i}d^3r M_i({\bf r})$,
  $\bar{e}_i=e_i({\bf r})/\int_{s_i}d^3r|e_i({\bf r})|^2$. So the
  $\bar{e}_i({\bf r})\propto e_i({\bf r})$ is a vector along the local magnetization
  density $M_i({\bf r})=n_\uparrow({\bf r})-n_\downarrow({\bf r})$, normalized by the
  total moment per sphere $M_i$, which is normalized by
  $\int_{s_i}d^3r\bar{e_i}({\bf r})\bar{e_i}({\bf r})=1$.  Here the indices $i,j$  refer to sites in the unit cell and $s_i$ is the muffin-tin radius of that site.

        The inverse $[{\bf D}^{0+-}({\bf q},\omega)]^{-1}$  at $\omega=0$ was shown to be related to the Heisenberg exchange interactions $J_{ij}({\bf q})$ in \cite{Kotani2008}.  By an inverse Bloch sum or discrete Fourier transform we extract the real space $J_{ij}^{{\bf 0 T}}$ between atoms $i$ in the unit cell at the origin and site $j$ in the unit cell at the lattice translation vector ${\bf T}$. Analyzing the corresponding
	Heisenberg Hamiltonian, we can then calculate the critical temperature in the mean-field or other approximations and also study the spin waves.
        The mean-field estimate of the critical temperature is typically used as a starting point for the self-consistent iterative process of the RPA scheme, first developed by Tyablikov \textit{et al.} \cite{Tyablikov1959} and Callen \cite{Callen63}. The mean-field estimate value is shown to be an upper limit of the RPA value by Rusz \textit{et al.} \cite{Rusz2005}.
        
	We note that 
	in this paper, we first concentrate on the collinear magnetic ordering (FM and AFM) within the unit cell of $I\bar{4}2d$ MnGeP$_2$ without including the effects of spin-orbit coupling (SOC). The Mn atoms are the dominant magnetic sites in the structure. When the SOC is neglected, it is indeed the above mentioned spin susceptibility which determines the fundamental magnetic linear response. The effect of spin-orbit coupling is primarily to lift some degeneracies at some specific {\bf k}-points and will be studied separately, but it does not affect
        Brillouin zone integrated quantities significantly, because only elements with fairly low atomic numbers are present in our system.  

        To obtain the SW dispersion within the Heisenberg model, we still only need the static and bare spin-susceptibility. However, Kotani and van Schilfgaarde \cite{Kotani2008} also proposed a way to obtain the interacting spin susceptibility under the assumption that the interaction is wavevector-independent and on-site diagonal. The resulting interacting and transverse dynamic spin-susceptibility but still within a rigid-spin within sites framework contains additional information on the lifetime of the spin-waves and is calculated from
        \begin{eqnarray}
          [D({\bf q},\omega)]^{-1}_{ij}&=&[D^0({\bf q},\omega)]^{-1}_{ij}+\frac{\omega}{M_i}\delta_{ij}\nonumber \\
          &&-M_i^{-1}\sum_kM_k[D^0({\bf q}=0,\omega)]^{-1}_{ki}\delta_{ij},\label{eq:DQ}
        \end{eqnarray}
        where $M_i$ is the magnetic moment of site $i$ and the ${\bf D}$
        and ${\bf D}^0$ are the site versions of the transverse spin susceptibility.
        
\subsection{N\'{e}el Temperature} \label{sec:tc}
We use both the mean-field and random phase approximations (RPA) to estimate the critical (N\'{e}el) temperature of antiferromagnetic MnGeP$_2$. For a system with various magnetic sites per unit cell, the mean-field critical temperature is calculated by considering the coupled system of equations,
\begin{equation}
	\left<s_i\right>= \frac{2}{3k_BT}\sum_{j} J_{ij} \left<s_j\right> ,
\end{equation}
where the $\left<s_j\right>$ vector is the average $z$-component of the unit vector, $\left<s_j^{\bm{R}}\right>$, pointing in the direction of the magnetic moment of the $j$ atom, located at site $\bm{R}$ within the unit cell. 
Thus, the mean-field critical temperature is obtained by solving the eigenvalue problem, 
\begin{equation}
	\sum_j\left[\frac{2}{3k_B}J_{ij} - T\delta_{ij}\right]\left<s_j\right> = 0,
\end{equation}
whose largest eigenvalue $T$ gives the critical temperature  $T_c$. Here
\begin{equation}
	J_{ij}=\sum_{\bm{T}}J_{ij}^{\bm{0T}}
\end{equation}
 and the sum excludes the on-site term, $J_{ii}^{\bm{00}}$ when $i=j$ \cite{Sasioglu2004}.    
 
The mean-field estimate of the critical temperature is typically used as a starting point for the self-consistent iterative process of the RPA scheme, \cite{Tyablikov1959,Callen63}. 
 
\subsection{Optical Dielectric Function}\label{sec:optics}
The optical dielectric function is calculated using both the \textit{independent (quasi)particle approximation} and the Bethe Salpeter equation in the Tamm-Damcoff approximation and with static $W$ \cite{Onida02} as implemented by Cunningham $et.$ $al.$ \cite{Cunningham23} in LMTO basis set within the {\sc Questaal} suite.
In the independent particle method, we find the imaginary part of the dielectric function, $\varepsilon_2$, by considering the direct transitions between the quasiparticle states $\psi_{n{\bf k}}$ with energies $\epsilon_{n{\bf k}}$ in the spin polarized $GW$:
\begin{equation}
  \begin {split}
    \varepsilon_2(\omega)&=\frac{8\pi^2}{\Omega \omega^2} \sum_{cv{\bf k}}(f_{v{\bf k}}-f_{c{\bf k}})\left|\bra{\psi_{v{\bf k}}}\hat{\bm{e}}\cdot \bm{v}\ket{\psi_{c{\bf k}}}\right|^2 \\ &\times\delta(\epsilon_{c{\bf k}}-\epsilon_{v{\bf k}}-\omega),
    \end{split}
\end{equation}
where $\Omega$ is the unit cell volume, $\hat{\bm{e}}$ the polarization and ${\bf v}$ the velocity operator and $f_{n{\bf k}}$ are Fermi occupation factors for the
conduction bands $c$ and valence bands $v$.
Note that here the velocity matrix elements enter rather than the momentum matrix elements. Since the QS$GW$ equation contains a non-local exchange correlation part, the velocity which is obtained from the commutator ${\bf v}=\dot{\bf r}=\frac{i}{\hbar}[{\bf r},H]$ includes a contribution from $[\tilde\Sigma_{xc}({\bf r},{\bf r}^\prime),{\bf r}]$. These are not trivial to evaluate as they involve $d\tilde\Sigma_{xc}/d{\bf k}$ and the current approach to evaluate them tends to overestimate the matrix elements somewhat \cite{Cunningham18}.

In the BSE in the Tamm-Dancoff approximation,\cite{Hanke78,Strinati1988,Rohlfing2000,Onida02} on the other hand, the dielectric  function is calculated from the eigenvalues and eigenvectors of a two-particle Hamiltonian which includes the electron-hole interactions kernel,
  \begin{equation}
    K(1234)=\delta(12)\delta(34)\bar{v}-\delta(13)\delta(24)W,
  \end{equation}
    with $\bar{v}$ the microscopic part of the bare Coulomb interaction $v$, \ie  omitting the long-range
    ${\bf G}=0$ part in a Fourier expansion.
    Here the numbers are a shorthand  for position, spin and time of each particle, $1=\{{\bf r}_1, \sigma_1,t_1\}$.
    Expanding this four-point quantity in the basis of one-particle eigenstates $\psi_{n{\bf k}}({\bf r})$,
    one obtains an effective two-particle Hamiltonian, given by
    \begin{eqnarray}
      H^{(2p)}_{n_1n_2{\bf k},n_1^\prime n_2^\prime{\bf k}^\prime}({\bf q})&=&\left(\epsilon_{n_2{\bf k}+{\bf q}}-\epsilon_{n_1{\bf k}}\right)\delta_{n_1n_1^\prime}\delta_{n_2n_2^\prime}\delta_{{\bf kk}^\prime}\nonumber\\
      &&-\left(f_{n_2{\bf k}+{\bf q}}-f_{n_1{\bf k}}\right) K_{n_1n_2{\bf k},n_1^\prime n_2^\prime{\bf k}^\prime}({\bf q})\nonumber \\
    \end{eqnarray}
    with $f_{n{\bf k}}$ the Fermi occupation function for band $n$ at ${\bf k}$.
    Effectively, this mixes up the above vertical transitions between valence and conduction band states $(\epsilon_{c{\bf k}}-\epsilon_{v{\bf k}})$ at different {\bf k}.  
    Diagonalizing this Hamiltonian in the Tamm-Dancoff approximation, where $n_1$ is restricted to be a valence state and $n_2$ a conduction band state,
    one obtains the exciton eigenvalues $E^\lambda({\bf q})$ and eigenvectors
    $A^\lambda_{n_1n_2{\bf k}}({\bf q})$.
    Introducing the shorthand $s=\{n_1n_2{\bf k}\}$, 
    the dielectric function is then given by
    \begin{eqnarray}
      \varepsilon_M(\omega)&=&1-\lim_{{\bf q}\rightarrow0}\frac{8\pi}{|{\bf q}|^2\Omega N_k}
        \sum_{ss^\prime}(f_{n_2^\prime{\bf k}^\prime+{\bf q}}-f_{n_1^\prime{\bf k}^\prime})\nonumber \\
        &&\rho_s({\bf q}) \sum_\lambda\frac{A^\lambda_s({\bf q})A^{\lambda*}_s({\bf q})}{E^\lambda({\bf q})-\omega\pm i\eta} \rho_{s^\prime}({\bf q})^*\label{epsmac}
    \end{eqnarray}
    with the matrix element
    \begin{equation}
      \rho_{n_1n_2{\bf k}}({\bf q})=\langle\psi_{n_2{\bf k}+{\bf q}}|e^{i{\bf q}\cdot{\bf r}}|\psi_{n_1{\bf k}}\rangle
    \end{equation}
    Here, we have assumed no spin polarization and a factor two for spin and $N_k$ is the number of {\bf k}-points in the Brillouin zone.
  The limit ${\bf q}\rightarrow0$ can be taken analytically, $e^{i{\bf q}\cdot{\bf r}}\approx 1+i{\bf q}\cdot{\bf r}$
    and then involves dipole matrix elements $\langle\psi_{n_2{\bf k}}| {\bf r}|\psi_{n_1{\bf k}}\rangle\cdot\hat{\bf q}$, where $\hat{\bf q}$ gives the
    direction along which we take the limit to zero and which corresponds to the polarization directions of the macroscopic tensor $\varepsilon_M(\omega)$.
    Finally, one converts the dipole matrix elements between Bloch states to velocity matrix elements divided by the band difference,
    $\langle \psi_{n_2{\bf k}}|[H,{\bf r}]|\psi_{n_1{\bf k}}\rangle=(\epsilon_{n_2{\bf k}}-\epsilon_{n_1{\bf k}})\langle\psi_{n_2{\bf k}}| {\bf r}|\psi_{n_1{\bf k}}\rangle$.

    For  a spin-polarized case our calculation includes only transitions between equal spins and neglects the coupling between opposite spin transitions.
    In principle these can arise from the exchange part of the kernel
    \begin {equation}
      \langle vc|K^x|v'c'\rangle=\int d{\bf x}d{\bf x}^\prime \psi_c^*({\bf x})\psi_v({\bf x})v({\bf r},{\bf r}^\prime)\psi_{c'}({\bf x}^\prime)\psi_{v'}^*({\bf x}^\prime),
    \end {equation}
    because the integration of spins included in $d{\bf x}$ requires the spins $\sigma_v=\sigma_c$ and  $\sigma_{v'}=\sigma_{c'}$ but the spins of primed and unprimed band pairs do not need to be equal.  In the non-spin polarized case, the  spatial parts of the up and down spin states are the same and the spin fine structure can be diagonalized separately for each eigenstate and leads to the well known splitting of spin singlet and triplet excitons
      \cite{Rohlfing2000}. In the spin polarized case, if the spin up and spin down states are sufficiently far away  and the exchange matrix elements of the kernel are small enough their interaction would only enter in second order perturbation theory and is therefore usually neglected.  Also, we note that
      as usual, the BSE calculation does not include indirect band to band transitions facilitated by electron-phonon coupling.

The implementation in terms of the LMTO
basis set involves the same product basis set also used for the QS$GW$ method following \cite{Cunningham18,Cunningham23}.

\section{Results}
The chalcopyrite crystal structure is shown in Fig.\ref{struc}. It belongs to the space-group number 122, $I\bar{4}2d$ or $D_{2d}^{12}$. The conventional body centered cell contains 16 atoms, but we use the primitive cell of 8 atoms. The labeling of the Mn atoms in this figure corresponds to the antiferromagnetic ordering as detailed later. The conventional unit cell's lattice parameters used in our calculations are $a= 5.453\textup{~\AA}$ and $c=10.780\textup{~\AA}$  with the ratio $c/a = 1.977$. The Wyckoff positions are: Ge, $4a$   $(0,1/2,1/4)$, Mn $4b$ $(1/2,0,1/4)$ and P $8d$ $(3/4,3u,3/8)$ with $u=0.255137$ the only internal structural parameter which is not fixed by symmetry. Because the system is body centered, we can use a 8 atom cell with ${\bf a}_1=(-1/2,1/2,1/2)$,  ${\bf a}_2=(1/2,-1/2,1/2)$,  ${\bf a}_3=(1/2,1/2,1/2)$ in terms of the lattice vectors of the conventional cell. The lattice structural parameters used here were obtained from the Materials Project (MP) \cite{Jain2013}, item  mp-1206503,\cite{mp-1206503} which were obtained by minimization of the total energy within the r2SCAN meta-GGA exchange correlation functional\cite{r2SCAN}. These lattice constants are about 4\% lower than the ones reported by Cho \etal\cite{Cho2004}, which are $a=5.655$ \AA, $c=11.269$ \AA\  or von Bardeleben \etal\cite{Bardeleben2024}, which are $a=5.655$ \AA,  $c=11.323$ \AA. 
The labeling of high symmetry points in the Brillouin zone follows the notation of the Bilbao Crystallographic Server, \url{https://www.cryst.ehu.es/}.

\subsection{Band Structures}

\begin{figure*}
	\includegraphics[width=15cm]{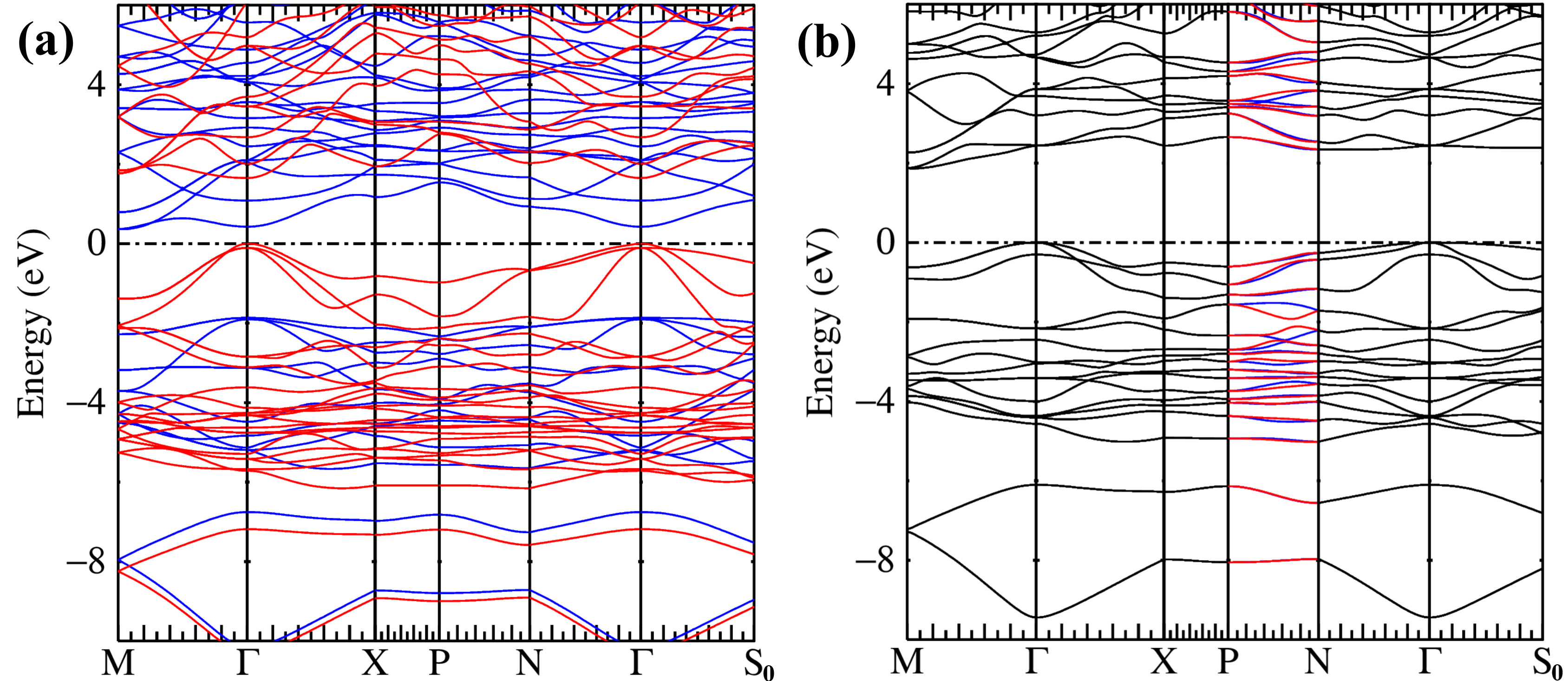}
	\caption{Band structure of MnGeP$_2$ obtained in the QS$GW^{BSE}$ approach, where the
		majority spin bands are shown in red and the minority spin bands are shown in blue. The Fermi energies are shifted to zero. (a) FM and (b) AFM. \label{fig-bnds}}
\end{figure*}

\begin{figure}
	\includegraphics[width=9cm]{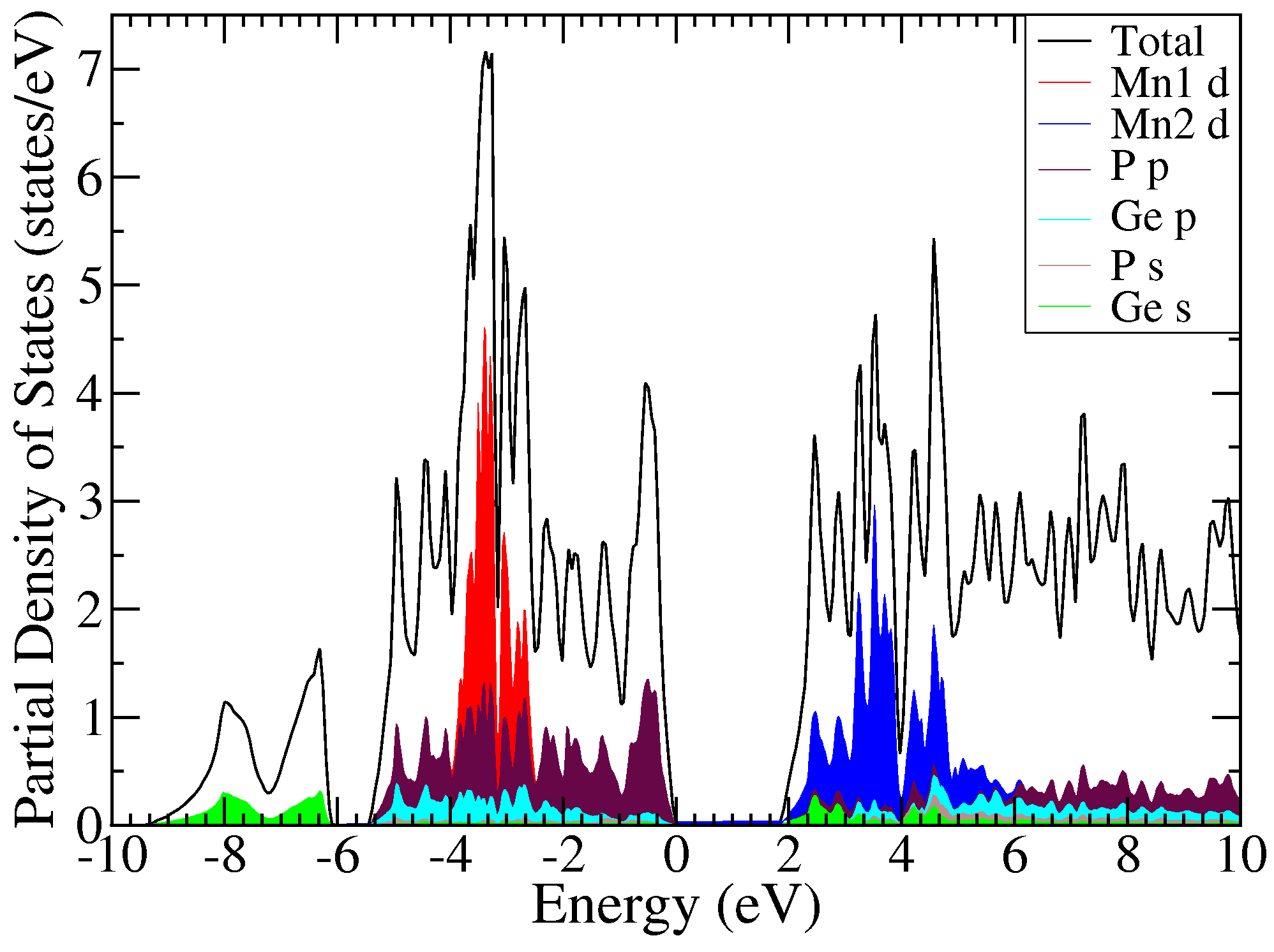}
	\caption{Total and partial densities of majority spin states (PDOS) in the QS$GW^{BSE}$ approach for AFM $I\bar{4}2d$ MnGeP$_2$. The partial contributions include the sum over all equivalent atoms of a given type but refer to partial wave contributions inside the spheres only, excluding those from the interstitial region and only showing the major contributions. 
    \label{fig-dos}}
\end{figure}

The Mn-$d$ state spin-polarization plays an essential role in making the band structure insulating. Without it, the partially filled $d$ states of Mn would lie at the Fermi level and there would be no band gap. Already in spin-polarized DFT in the GGA  we find that spin up and spin down states of Mn are fairly well
separated. In the ferromagnetic case, however, the band structure is still metallic. In order to open a gap, we can add a Hubbard $U$ which increases the exchange splitting of the Mn-$d$. In the antiferromagnetic case, a gap is already present at the spin-polarized DFT  level. We here use a spherically averaged GGA+$U$ approach\cite{Dudarev98}  in which only an effective $\bar{U}=U-J$ enters where $J$ is the exchange term. Note that this is justified because Mn corresponds to a half-filled shell. Adding $U$ reduces the hybridization of the occupied Mn-$d$ states with the P-$p$ states in the occupied bands, which then dominate the valence band maximum (VBM). Likewise, the empty  Mn-$d$ states are interacting less with the Ge-$s$ states, which becomes the conduction band minimum (CBM). The  gap opens up depending on the  choice of $U$. Starting from the Green's function $G^0$ of this  DFT$+U$ band structure,
we then calculate the self-energy $\tilde\Sigma^0_{xc}$. This first step  $\tilde\Sigma^0_{xc}$ is then added to the DFT Hamiltonian without Hubbard $U$ and its eigenvalues and eigenstates provide the input for the next $GW$ step. From here on,  we let the self-energy $\tilde\Sigma_{xc}$ evolve till its change from the previous iteration falls below a tolerance of $1\times10^{-5}$ Ry.  Essentially the self-energy takes over the role of the initial Hubbard $\bar{U}$. Since the initial $\bar{U}$ is used only to provide a reasonable starting band structure, its value is not crucial and the final results should be independent of this choice. We used a value of $\bar{U}=5.4$ eV.
In fact, it turns out that even  without $U$ the same final QS$GW$ results
are obtained. 
Using the same approach we calculate both the FM and AFM band structures and we also use both the original QS$GW^{RPA}$ method  and the improved QS$GW^{BSE}$ methods as well as the GGA+U, and spin-polarized GGA methods.  The values reported as $G^0W^0$ are the first iteration results of the QS$GW^{RPA}$  cycle and do already include off-diagonal matrix elements of $\tilde\Sigma_{xc}$ and a re-diagonalization of the $H^0$ in contrast to the usual procedure of just including the diagonal corrections to the eigenvalue by perturbation theory.  
The final QS$GW^{BSE}$ band structures obtained in this way for the FM and AFM cases are shown in Fig. \ref{fig-bnds} and the band gaps are listed in Table \ref{bands}.

\begin{table}[h]
    \centering
    \caption{Band gaps in ferromagnetic and antiferromagnetic $I\bar{4}2d$ MnGeP$_2$. See supplemental information \cite{supinfo} for the associated band plots.}
    \begin{ruledtabular}
      \begin{tabular}{cccccc}
        \multicolumn{6}{c}{FM} \\ \hline
 transition & \multicolumn{5}{c}{gap (eV)}\\ \hline
        &GGA&GGA+$U$&$G^0W^0$&QS$GW^{RPA}$&QS$GW^{BSE}$\\ \hline
        $\Gamma\uparrow-\Gamma\downarrow$ &$-$1.028 &0.086 &0.948 &0.634 &0.426 \\
        $\Gamma\uparrow-\Gamma\uparrow$ &0.707 &1.042 &1.659 &1.799 &1.653 \\
        $\Gamma\downarrow-\Gamma\downarrow$ &1.486 &1.597 &2.276 &2.476 &2.293 \\ 
        $\Gamma\uparrow-M\downarrow$ &$-$1.199 &$-$0.143 &0.931 &0.571 &0.366 \\ \hline
        \multicolumn{6}{c}{AFM} \\ \hline
        $\Gamma-M$ &0.592 &1.129 &1.875 &1.939 &1.869 \\
        $\Gamma-\Gamma$ &0.898 &1.717 &2.384 &2.490 &2.440 \\
      \end{tabular}
    \end{ruledtabular}
    \label{bands}
    \end{table}

For the FM case,  the top of the valence band has majority spin while the lowest conduction bands have minority spin. The majority spin VBM and the CBM of minority spin are located at 
$\Gamma$ and $M$ respectively implying an  indirect  and  opposite spin gap. The gaps are all slightly reduced in the QS$GW^{BSE}$ case,  which is shown in Fig\ref{fig-bnds}(a), compared to QS$GW^{RPA}$. The band structures in GGA, GGA+$U$,  $GW^{RPA}$ first iteration and QS$GW^{RPA}$ are given  in supplementary material.\cite{supinfo}  Note that in the FM case, the GGA band structure is metallic as the minority spin Mn-$d$ set of bands are still overlapping with the bands identified as VBM and CBM in the QS$GW^{RPA}$ cases. The transition energies quoted in the table \ref{bands} are for the corresponding states which become VBM and CBM for each spin in the QS$GW^{RPA}$ case. 

The band structure for the AFM case in which the two atoms in the unit cell have opposite magnetic moments and which is found to have lower total energy within the GGA and GGA+$U$ approximation, is shown in Fig. \ref{fig-bnds}(b). The lowest band gap is again indirect between CBM at $M$ and VBM at $\Gamma$ but the difference is rather small from the direct gap at $\Gamma$.

The total density of states and its decomposition into partial orbital contributions is shown in Fig. \ref{fig-dos} for the AFM case in the QS$GW$ approximation. We can see that the Mn-$3d$ states occur around $-4$ to $-2$ eV and spread from 2 to 5.5 eV respectively but the majority spin (occupied) and minority spin bands (empty) occur on  the different Mn although overall the total DOS  of minority and majority spin are the same as required for an antiferromagnet. The top valence bands have primarily P-$p$ character. The lowest bands near $-8$ eV have mixed P-$3s$  P-$3p$ character while the mostly P-$3s$ bands occur near -29 eV and are not shown here. They lie close to the Ge-$3p$ semi core states. 

\begin{figure*}
	\includegraphics[width=15cm]{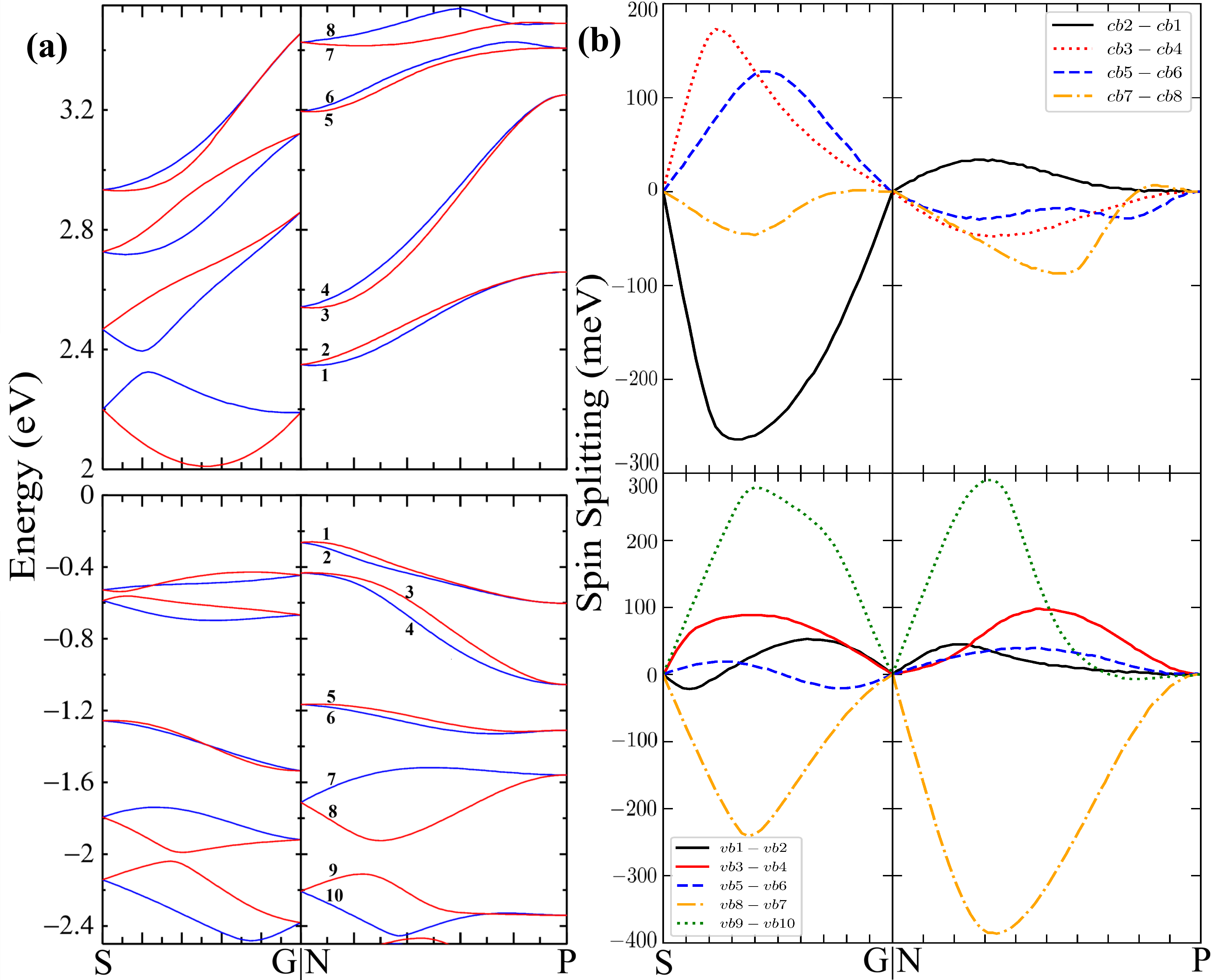}
	\caption{(a) Conduction and valence bands obtained in the QS$GW^{BSE}$ level along the symmetry lines where major splittings occur: S=$(1/2-\eta,1/2-\eta,1/2)$; G=$(0,1/2-\zeta,1/2)$; N=$(1/4,1/4,1/4)$; P=$(0,1/2,1/4)$ in Cartesian coordinates with $\eta=\frac{1}{4}\left(1+\frac{a^2}{c^2}\right)$ and $\zeta=\frac{a^2}{2c^2}$. All $k_z$ values are scaled by $\frac{c}{a}$. The majority spin bands are shown in red and the minority spin bands are shown in blue. (b) Differences of majority minus the minority energies in successive bands. The labels marking the differences utilize the numbering in (a).
		\label{fig-spin-splitting}}
\end{figure*}

Although the band plots in the AFM case show the spin up and spin down bands are degenerate along several symmetry lines, we see a splitting of the bands along $PN$. This is a signature of altermagnetism. Indeed, the space-group $I\bar{4}2d$ supports altermagnetism \cite{Smejkal2022}. The symmetry elements that relate the spin-up and spin-down Mn$_1$ and Mn$_2$ are generated by the 2-fold rotation about the $x$ or $y$ axis. They split the group $D_{2d}$ into the subgroup $S_4$ and its coset $C_{2x}S_4$
containing the elements $\{C_{2x},C_{2y},\sigma_{(110)},\sigma_{(1\bar{1}0)}\}$. For
any {\bf k} in one of the diagonal mirror planes, the mirror operation belongs to the little group of {\bf k} and the spin degeneracy is protected \cite{Smejkal2022}. This includes the bands along $\Gamma M$ and $\Gamma N$. Likewise along $\Gamma X$, the
$C_{2x}$ operation protects the spin-degeneracy while along $XP$ it is a combination of time reversal
${\cal T}$ with a two fold rotation $C_{2y}$ that protects the degeneracy. However along  $SG$ and $NP$ rather sizable spin splittings are observed. Following the approach of Yuan \etal \cite{YuanLD2020}, the splittings as a function of ${\bf k}$ along the $SG$ and $NP$ lines are plotted separately for a few of the bands in Fig. \ref{fig-spin-splitting}. While the spin splitting is zero along the symmetry lines shown near the CBM at $M$ and the VBM at $\Gamma$ spin splitting
should occur at neighboring points in arbitrary directions and thus the
altermagnetic spin splitting could have interesting effects on transport
in carrier doped MnGeP$_2$. 

Finally, we discuss the effects of spin-orbit coupling. We add spin-orbit coupling (SOC) to the QS$GW$ calculation in the manner described by Brivio \etal\cite{Brivio2014}. The $L_zS_z$ SOC term maintains the spin character of the bands and can be readily included in the QS$GW$ cycle. The $\frac{1}{2}(L_+S_-+L_-S_+)$
terms which mix the spins are added perturbatively at the level of the eigenvalues, while maintaining the spin character. The full band structure obtained in this way within QS$GW^{RPA}$ is given in Supplemental Material\cite{supinfo}. We find that at $\Gamma$ the doubly degenerate VBM undergoes a SOC splitting of 25.8 meV. These states remain spin degenerate.
As we move slightly away from $\Gamma$ we can now observe small spin-splittings resulting from SOC even along the lines along which spin degeneracy was preserved without SOC. These however are of the order a few 10 meV only. The spin-splittings along the lines PN which are related to the altermagnetic character are larger and stay almost the same as without SOC. 
\subsection{Optical Properties}

\begin{figure}
	\includegraphics[width=9cm]{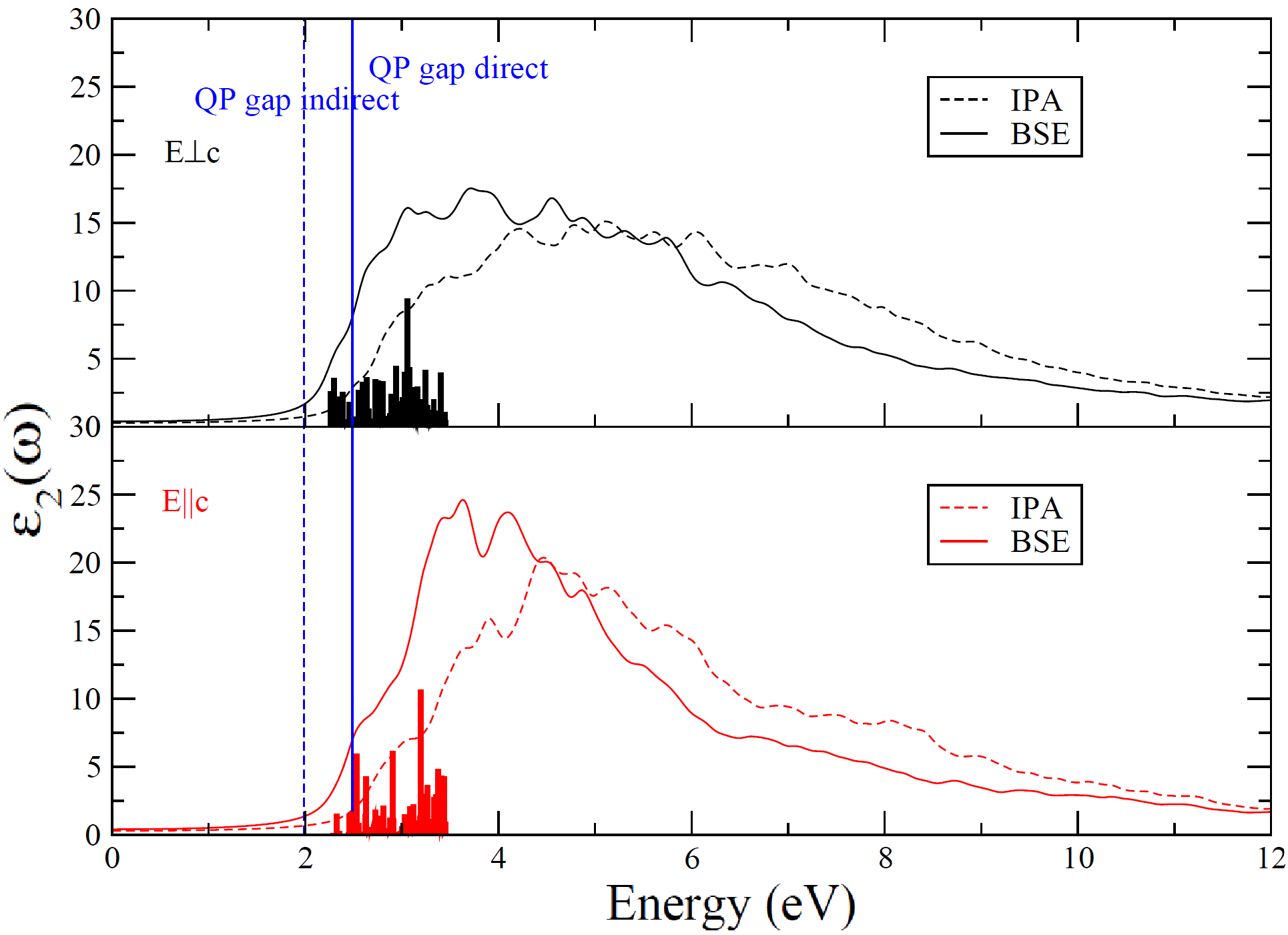}
	\caption{$\epsilon_2 (\omega)$ obtained within the Bethe-Salpeter and independent particle approximation schemes applied in the $GW$ level. The vertical bars show the individual exciton (two-particle Hamiltonian) eigenvalues with height proportional their oscillator strengths in the low energy region up to +1 eV above the direct quasiparticle gap.
	  \label{fig-bse}}
\end{figure}

Absorption or reflectivity or spectroscopic ellipsometry in the visible to ultraviolet range provide a convenient way to probe the band structure via interband transitions. To facilitate future comparisons to experimental work of the interband related optical properties, 
we have calculated the optical dielectric function both in the independent particle approximation and in the BSE approach, which includes local-field and electron-hole interaction effects.
Because the chalcopyrite structure has tetragonal symmetry, the macroscopic dielectric function has two independent components, parallel or perpendicular to the
{\bf c}-axis. Optical properties are studied in the AFM case since we find the AFM configuration to be the lowest energy. The results, shown in Fig. \ref{fig-bse}, are based on the QS$GW^{BSE}$ band structure. 

 Our calculation includes a fixed broadening factor which simulates both the lifetime and the spread in {\bf k}-space around the Brillouin zone sampling {\bf k} points. We use a $5\times5\times5$ mesh and a broadening parameter of 0.005 Rydberg. The BSE results are also sensitive to the number of valence and conduction bands included in the transitions. Here we include 22 valence and 16 conduction bands. This includes the upper valence band set derived primarily from the 12 P-$3p$ orbitals and the 10 Mn-$3d\uparrow$ orbitals in the AFM case, which extend down to about $-5$ eV from the VBM and up to about 8 eV in the conduction band. This should be adequate to obtain $\varepsilon_2(\omega)$ converged up to about 13 eV.

 We can see that the BSE shifts the peaks in $\varepsilon_2(\omega)$ to lower energy and changes the shape of the spectrum. higher energy peaks become weaker and lower energy peaks in the continuum become stronger. 
 In other words, the oscillator strength is red-shifted.  Examining the eigenvalues of the effective two-particle Hamiltonian, and their relative oscillator strengths, we can see more detail of the lowest excitations without the broadening effects. These are plotted as narrow vertical bars on the same plot but with an arbitrarily adjusted overall scale factor.
We can now see that there are in fact bound state excitons below the direct gap of 2.44 eV. We note, however that these all are still lying above the lowest indirect gap of 1.87 eV. 
Our present calculation does not include indirect contributions which require inclusion of electron-phonon mediated second order processes and are therefore expected to be significantly weaker. We anticipate that there will be also bound state excitons associated with the indirect gap and with similar exciton binding energies. While the optical dielectric function is well converged in the continuum region in terms of {\bf k}-point sampling and number of bands included to give reliable results at the 0.1 eV scale, the binding energy of the excitons
at meV scale is more sensitive to the {\bf k}-point sampling and requires a finer sampling to be accurate than we can here afford. We have found in previous works \cite{Dadkhah2023,Dernek-zgn2024,Dadkhah2024} that for bound excitons of the Wannier type, one can use a smaller number of bands but needs an extrapolation in terms of the density of {\bf k}-points near the {\bf k}-point where the direct gap occurs. Even then, exciton binding energies tend to be severely overestimated except in wide band gap systems\cite{Dadkhah2023} or 2D materials \cite{Dadkhah2024} because the BSE method, as presently implemented, does not include lattice contributions in the screening. 
We therefore refrain from a detailed discussion of the exciton binding energies or exciton spectrum. Here we point out that there are indeed excitons and that on the larger energy scale in the continuum the electron-hole interaction effects are significant. We can also see significant anisotropy between the ${\bf E}\parallel {\bf c}$ and ${\bf E}\perp{\bf c}$ spectra. 

\subsection{Magnetic Properties}
\begin{figure}
  \includegraphics[width=9cm]{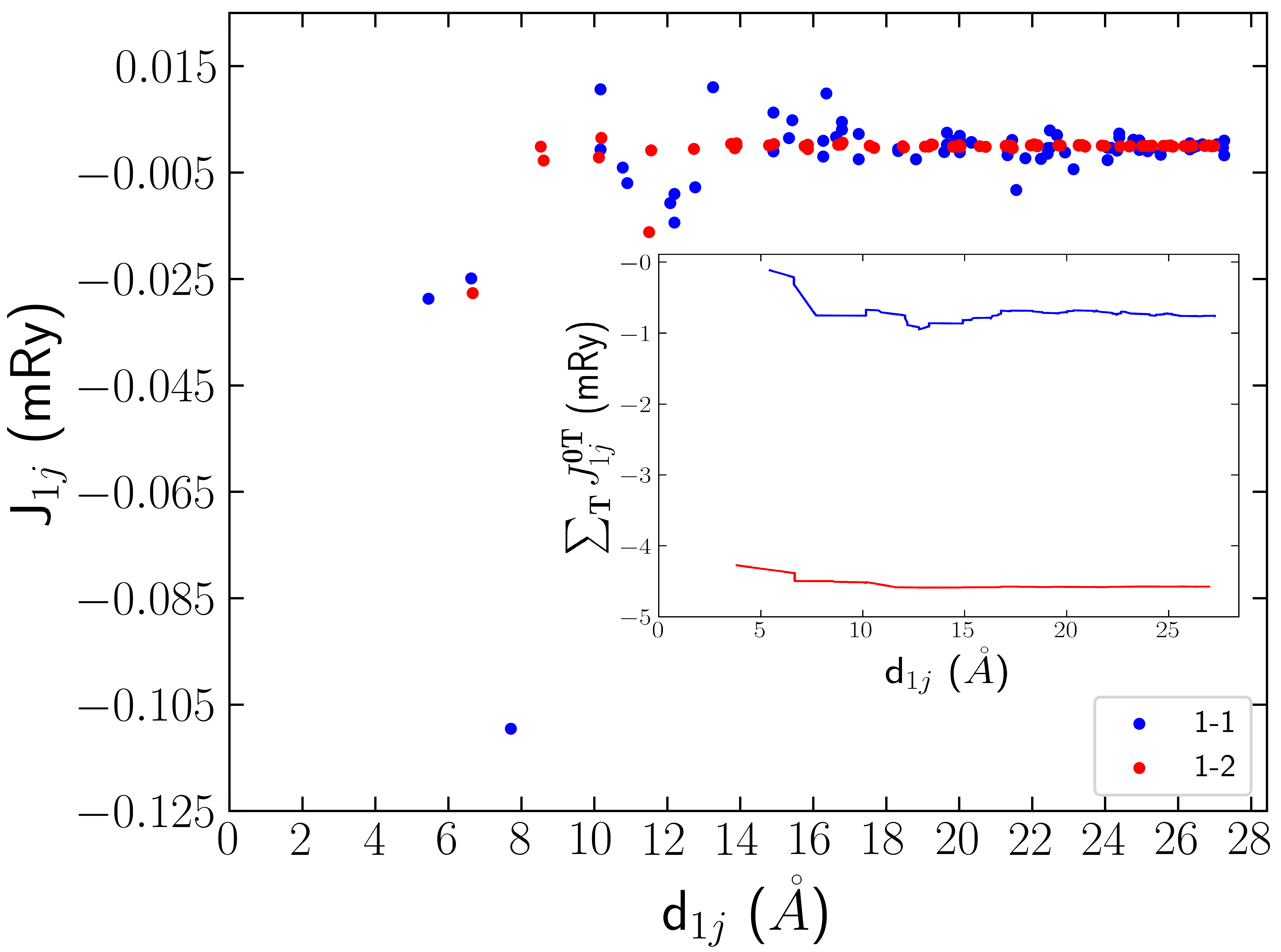}
  \caption{Exchange interactions $J_{11}^{\bf T}$ in blue and $J_{12}^{\bf T}$ in red as function
    of distance  $d_{1j}=|\tau_{1} - \tau_{j} - \mathbf {T}|$ from the AFM ordering using QS$GW^{BSE}$ states. The nearest neighbor $J_{12}^{\bf (0,0,0)}$
    is left off as it is at least 10-fold larger in absolute value from the other ones. The inset shows the cumulative sums $\sum_{\bf T}J_{11}^{{\bf 0 T}}$ in blue and $\sum_{\bf T}J_{12}^{{\bf 0 T}}$ in red, where the largest $J_{12}^{\bf (0,0,0)}$ interaction is also included. See Table \ref{exchange-interactions} for some of the major interactions and the net sums obtained. \label{figJd} }
\end{figure}

Having established the electronic band structure for both FM and AFM cases, we now turn to the question of magnetic ordering in the perfect crystal. The calculations of magnetic properties in this section is based on the QS$GW^{BSE}$ underlying electronic structure. Corresponding results based on GGA and on QS$GW^{RPA}$ are given in Supplementary Information.\cite{supinfo} We note that the main exchange interaction based on the QS$GW^{RPA}$ band structure is about 5 \% smaller which is consistent with the band gaps being slightly higher. Using the GGA band structure overestimates the exchange interactions by about a factor two.

We label the exchange interactions $J_{ij}^{\bf T}$ between atom number $i$ in the unit cell centered at the origin and $j$ in the unit cell with lattice vector
${\bf T}=n_1a\hat{\bf x}+n_2a\hat{\bf y}+n_3c\hat{\bf z}$ denoted as $(n_1,n_2,n_3)$ in Table \ref{exchange-interactions}. In other words, the lattice vectors here are given in reduced coordinates of the conventional cell.
The table lists just a few of the larger interactions and the total cumulative sum. The dependence of these exchange interactions on distance is shown in Fig. \ref{figJd}. We can see that apart from the nearest neighbor interactions they quickly converge to almost zero with some minor oscillations.
We find that the major interaction is  $J_{12}^{(0,0,0)}$, which is the nearest neighbor interaction between the Mn$_1$ and Mn$_2$ sites in the same primitive cell, as shown in Fig. \ref{struc}(a)), and is negative for both FM and AFM reference calculations. This shows clearly that pure MnGeP$_2$ is predicted to be AFM confirming various previous DFT studies\cite{ZhaoYJ2001,Mahadevan2003}.
On the other hand we find the exchange interactions in the (001) planes to be weakly antiferromagnetic, $J_{11}^{(\pm100)}=J_{11}^{(0\pm1 0)}=-0.028$ mRy and $J_{11}^{(110)}=-0.109$ mRy and also the $J_{11}^{\pm1/2,\pm1/2,\pm1/2}=-0.0249$ which connect atoms by primitive lattice vectors. 
Besides the clear antiferromagnetic ordering between adjacent (001) planes, this suggests there could be a further antiferromagnetic checkerboard ordering pattern in these planes but these interactions are one to two orders of magnitude smaller  and the stronger interactions in the [110] direction and the weaker in the [100] directions are  frustrated with respect to each other. The antiferromagnetic coupling by the [1/2,1/2,1/2] type interactions on the other hand are compatible with the AFM ordering along [001] planes.

\begin{table}[h]
	\centering
	\caption{Exchange interactions based on FM or AFM reference state; $i$ and $j$ label the magnetic atom site numbers in the primitive cell, and ${\bf T}$ gives the lattice vector in units of the conventional cell lattice vectors, n${^*}$ is the number of equivalent neighbors in the star, $J_{ij}^{{\bf T}}$ is the exchange interaction in mRy, and $\sum_{\bf T}J_{ij}^{{\bf 0 T}}$ is the cumulative sum up to $|\tau_{i} - \tau_{j} - \mathbf {T} | < r_{cut} = 5a\approx27$\AA, where "all" stands for the net sum obtained. $\tau_2 - \tau_1=(0,-0.5,0.25)$ in reduced coordinates of the conventional cell.}
	\begin{ruledtabular}
		\begin{tabular}{c c c c c c c}
			Ordering & i & j & n${^*}$ & $\bf T$ & $J_{ij}^{{\bf 0 T}}$ (mRy)& $\sum_{\bf T}J_{ij}^{{\bf 0 T}}$ (mRy)\\ [1ex]
			\hline
			FM & 1 & 1 & 4 & (1,0,0) & 0.0072 & -0.029 \\ [1ex]
			FM & 1&	 1 & 8	& (1/2,1/2,1/2) & 0.0064 & 0.080 \\[1ex]
			FM & 1 & 1 & 4 & (1,1,0) & -0.0748 & -0.220 \\ [1ex]
			FM & 1 & 1 & 2 & (0,0,1) & -0.0239 & -0.275 \\ [1ex]
			FM & 1 & 1 &   &  all      &         & -0.203 \\ [1ex]
			FM & 1 & 2 & 4 & (0,0,0) & -1.0211 & -4.084 \\ [1ex]
			FM & 1 & 2 & 8 & (1,0,0) & -0.0284 & -4.312 \\ [1ex]
			FM & 1 & 2 &   &  all      &         & -4.555 \\ [1ex]
			\hline
			AFM & 1 & 1 & 4 & (1,0,0) & -0.0287 & -0.115 \\ [1ex]
			AFM & 1 & 1 & 8 & (1/2,1/2,1/2) & -0.0249 & -0.314 \\ [1ex]
			AFM & 1 & 1 & 4 & (1,1,0) & -0.1095 & -0.752 \\ [1ex]
			AFM & 1 & 1 &   &  all      &         & -0.761\\ [1ex]
			AFM & 1 & 2 & 4 & (0,0,0) & -1.0687 & -4.275 \\ [1ex]
			AFM & 1 & 2 & 8 & (1,0,0) & -0.0277 & -4.496 \\ [1ex]
			AFM & 1 & 2 & 4 & (0,1,0) & -0.0027 & -4.507 \\ [1ex]
			AFM & 1 & 2 & 8 & (1,0,1) & -0.0021 & -4.525 \\ [1ex]
            AFM & 1 & 2 & 4 & (3/2,1/2,1/2)& -0.0162 & -4.577 \\ [1ex]
			AFM & 1 & 2 &   &  all      &         & -4.575 \\ [1ex]
		\end{tabular}
	\end{ruledtabular}
	\label{exchange-interactions}
\end{table}

The primitive AFM cell calculation yielded converged magnetic moments of Mn$_1$ and Mn$_2$ to be $\pm 4.576 \mu_B$, with zero net magnetic moment, whereas in the FM calculation, each site's magnetic moment is found to be $4.812 \mu_B$, with a net moment of $10 \mu_B$ including the small contributions from the Ge-$d$ and P-$d$ orbitals and interstitial contributions. These magnetic moments correspond closely to $S=5/2$ as expected for Mn$^{2+}$. 

We can see that the dominant exchange interaction is between the closest site 1 and site 2 Mn and is about $-1$ mRy with a slightly higher absolute value in the AFM case than in the FM reference case.  All other interactions are at least  an  order of magnitude smaller and most are two to three orders smaller.   Each Mn$_1$ has 4 nearest neighbor Mn$_2$ atoms in the primitive cell in the adjacent layers in the c-direction. If we ignore all other interactions, the mean field $T_c$ would be $(2/3)NJ_{12}^{nn}$, with $N$ the number of nearest neighbors within a classical Heisenberg model with unit vectors as spins and gives an estimated mean field temperature of 450 K. Correcting this by a quantum correction factor of $S(S+1)/S^2\approx1.4$ we would obtain $T_N^{MF}\approx630$ K. This is an overestimate because of the mean field approximation. We may compare to the case of MnO in \cite{Kotani2008}. In that case the authors found that the $T_N$ already agreed reasonably well with experiment even without the quantum correction. 
 
A more accurate calculation of the critical temperature is obtained
using the approach described in Sec. \ref{sec:tc} and is used as starting point for the Tyablikov method. The Tyablikov method is sensitive to the {\bf k}-point sampling and may converge to different fixed points. Using up to a $15\times15\times15$ mesh we obtain a converged N\'eel temperature of 401 K in the mean field approximation and 173$\pm$2 K in the Tyablikov method.  
 Multiplying the RPA estimate by a quantum correction as mentioned before would yield about 242 K. At present we have no experimental data on AFM MnGeP$_2$ as all samples yield ferromagnetic behavior for reasons to be discussed in later sections. 

We next turn our attention to the spin waves which, in principle, could provide a more direct test of the exchange interactions, although unfortunately no such experimental data are presently available. Fig. \ref{fig-chipm} shows the SW energies $\omega({\bf q})$ of the pure crystal obtained from
\begin{equation}
  \det{\omega({\bf q})\delta_{ij}\frac{1}{S_i}-\frac{1}{S_i}\tilde J_{ij}^{\cal H}({\bf q})\frac{1}{S_j}}=0
\end{equation}
with
\begin{equation}
  \tilde J_{ij}^{\cal H}({\bf q})=J_{ij}^{\cal H}({\bf q})-\delta_{ij}\frac{1}{S_i}\left(\sum_kJ_{ik}^{\cal H}({\bf q}=0)S_k\right)
\end{equation}
with $2S_i=M_i$ and the $J^{\cal H}_{ij}({\bf q})=[D^0({\bf q},\omega=0)]^{-1}$ of Eq. (\ref{eq:DQ}) apart from the on-site terms.
Within the simplified model of only one nearest neighbor interaction $J_{12}$ we can obtain the SW energy analytically as
\begin{equation}
  \omega({\bf q})=|J_{12}|\sqrt{z^2-\left(\sum_{\bf d} e^{i{\bf q}\cdot{\bf d}}\right)^2}
\end{equation}
with $z$ the number of nearest neighbors, which in our case is 4 and
the sum over neighbors amounts to
\begin{equation}
  \sum_{\bf d} e^{i{\bf q}\cdot{\bf d}}=2\cos{(k_xa/2)}e^{ik_zc/4}+2\cos{(k_ya/2)}e^{-ik_zc/4}
\end{equation}
which at $k_X=(0,\pi/a,0)$ amounts to $|J_{12}|2\sqrt{3}$ or about 50 meV for $\omega({\bf q})$. At  $k_M=(0,0,2\pi/c)$ the SW energy is $4|J_{12}|$ or 58 meV and at $k_N=(\pi/2a,\pi/2a,\pi/c)$, it is $|J_{12}|2\sqrt{2}$ or 41 meV. These values are in excellent agreement with those shown in Fig.\ref{fig-chipm}, thus validating the model with only one dominant exchange interaction.    A measurement of the SW dispersion could directly confirm our exchange interaction estimates.

Besides the SW dispersion relation shown as green dashed lines in Fig.\ref{fig-chipm} this figure show the imaginary part of the trace of the $D({\bf q},\omega)$ which represents the site averaged dynamic transverse spin susceptibility. The magnitude of this quantity is shown as a color scale as function of {\bf q} and $\omega$. We can see that its maximum nicely follows the SW energy as expected. In principle,  the width of  
${\rm Im}\left\{{\rm Tr}[{\bf D}({\bf q},\omega)]\right\}$ is the inverse of the lifetime of the collective SW excitation, which, physically arises from decay into spin-flip excitations. While in an altermagnet we indeed have small energy spin splittings in the bands these can only become excited if we first have electron hole pairs excited to the band edges at finite temperature, which is not included here. The smallest spin flip excitations allowed in our calculation are across  the band gap and thus provide only negligible width because the electronic band gap is much larger than the SW energy maximum.

\begin{figure}
	\includegraphics[width=9cm]{Figs/sw_heatmap_plot.png}
	\caption{${\rm Tr}\left\{{\rm Im}[\chi^{+-}(\omega)]\right\}$ as color scale and spin wave (SW) dispersion, $\omega(\textbf{q})$ (yellow dashed line) for AFM MnGeP$_2$.	
	\label{fig-chipm}}
\end{figure}

\subsection{Effects of Carrier Doping}

Past studies (\cite{Sato2005}, \cite {Cho2004}, \cite{Bardeleben2024}) on synthesized MnGeP$_2$ found ferromagnetic hysteresis and resonance. To explain the possible source of ferromagnetism, we first analyze the effect of carrier mediated exchange interactions on the dominant antiferromagnetic exchange interaction, by introducing a simple model for carrier doping. We basically assume a rigid band structure but change the occupation of the valence or conduction bands by adding a positive or a negative homogeneous background charge density to the structure. Introducing a negative background charge simulates the effects of hole doping, since the charge neutrality requires fewer valance electrons. Whereas introducing a positive background simulates electron doping.
The resulting dominant interactions under each type of doping are listed in table \ref{carrier_doping}. 

We can see that both hole or electron doping lowers the absolute value of the relevant exchange interaction and the decrease increases with the doping concentration but is not simply proportional to it. It appears that electron doping is more effective at lowering the antiferromagnetic exchange, but even for the high concentration of dopants simulated here, like 1 electron or hole per unit cell, or about 5.4$\times 10^{21}$ cm$^{-3}$, the effect is less than 5\%. 

\begin{table}[]
	\centering
	\caption{Effect of carrier doping by introducing a homogeneous background charge }\begin{ruledtabular}
	\begin{tabular}{c c}
	  Charge (e)& $J_{12}$ (mRy) \\[0.5ex]
                 -1 & -0.99976 \\ [1ex]
		-0.5 & -1.00259 \\ [1ex]
		-0.1 & -1.01415 \\ [1ex]
		0 & -1.01882 \\ [1ex]
		0.1 & -1.00876 \\ [1ex]
		0.5 & -0.98269 \\ [1ex]
		1 & -0.95391 \\ [1ex]
	\end{tabular}
        \end{ruledtabular}
	\label{carrier_doping}
\end{table}

We conclude that the carrier mediated interactions cannot explain the observed ferromagnetism in MnGeP$_2$. 

\subsection{Mn$_{\rm Ge}$ anti-site defects}

\begin{figure}
	\includegraphics[width=7cm]{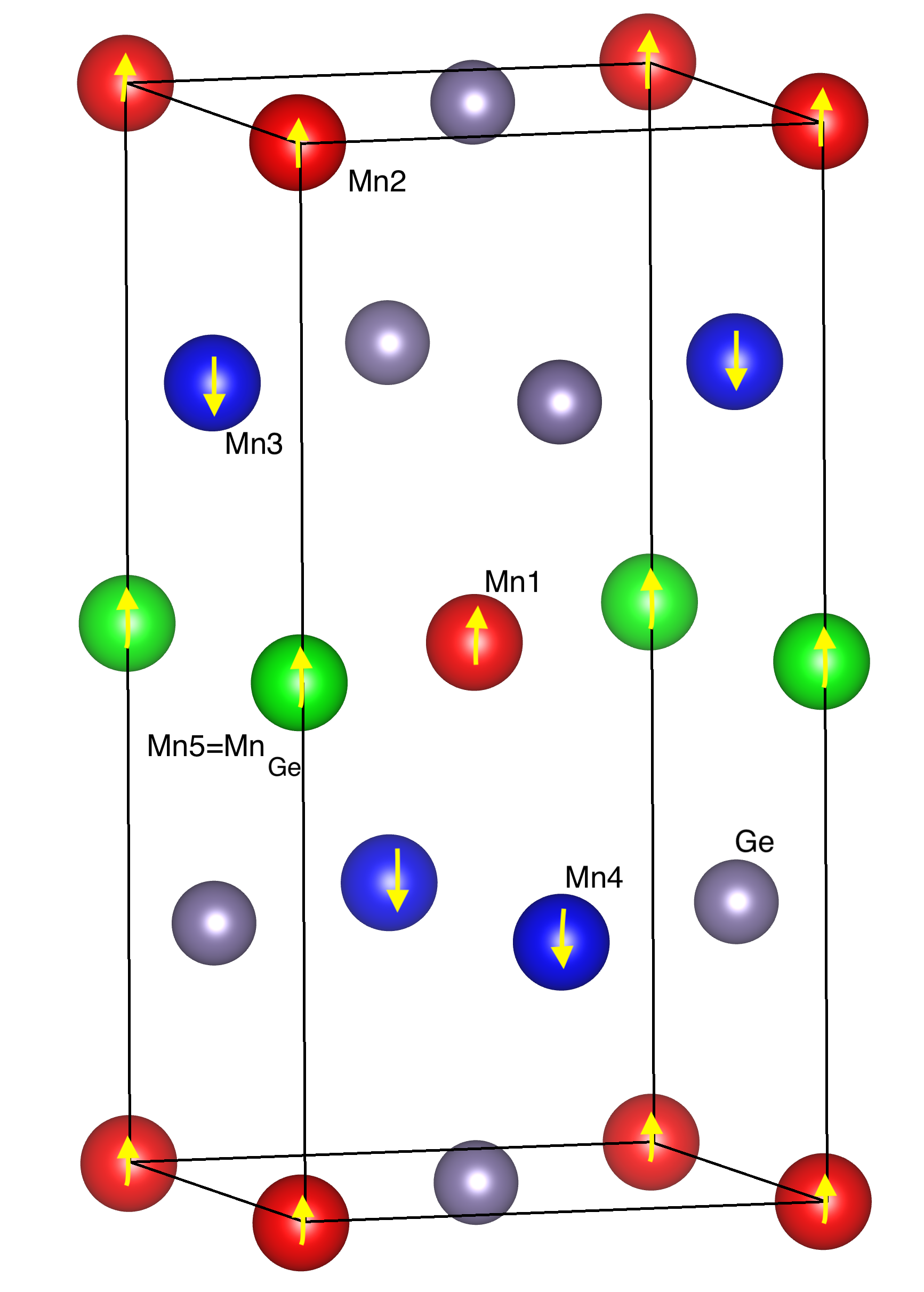}
	\caption{Labeling of the Mn atoms and assumed spin configuration in the 16 atom conventional cell with Mn$_{\rm Ge}$ antisite. Mn atoms with spin-up are shown in red and Mn atoms with spin-down are shown in blue. Ge are gray and P atoms are not shown. The green sphere is the Mn$_{\rm Ge}$ antisites is assigned to up spin. The Mn$_4$ atoms nearest to Mn$_5$ correspond to ${\bf T}=(0,0,-1)$.
         \label{fig-supercell-16}}
\end{figure}

As a second possible source of ferromagnetic interactions, we consider Mn$_{\rm Ge}$ antisite defects based on the suggestions of the work of Mahadevan and Zunger \cite{Mahadevan2003}. We consider several  models. In the first one, we use the 16 atom conventional cell and replace one Ge by a Mn. Effectively this changes the stoichiometry and adds a full layer consisting of MnP. While bulk MnP has a different crystal structure, this model nonetheless simulates in some approximate sense a MnP precipitate but maintaining the tetrahedrally bonded zincblende structure. Note, that we are not claiming the existence of such a monolayer type precipitate. Rather, we are exploring the qualitative effect of adding Mn on Ge sites. In this model, the mean-field critical temperature estimate is 560.4 K and the RPA estimate is 390.9 K. See Fig. \ref{fig-supercell-16} and table \ref{tab:J16} for the detailed tabulations of the exchange interactions between the antisite Mn and the other Mn within the structure. 

\begin{table}
  \caption{Exchange interactions involving Mn$_{\rm Ge}$ in 16 atom cell.
    The atom labeling is shown in Fig. \ref{fig-supercell-16}. Mn$_1$ ($\bm{\tau}_1=(1/2,1/2,0)$ in reduced coordinates) and  Mn$_3$ ($\bm{\tau}_3=(0,0,1/2)$) are $\uparrow$-spin, Mn$_2$ ($\bm{\tau}_2=(1/2,0,1/4)$) and Mn$_4$  ($\bm{\tau}_4=(0,1/2,3/4)$) are $\downarrow$-spin and Mn$_5$ is the Mn$_{\rm Ge}$ with $\uparrow$-spin; ${\bf T}$ gives lattice translation vectors, $|{\bf R}|=|\bm{\tau}_j-\bm{\tau}_i+{\bf T}|$ in \AA, $N_j$ the number of equivalent neighbors and $J_{ij}^{{\bf T}}$ gives the exchange interaction in mRy. \label{tab:J16}}
    \begin{ruledtabular}
      \begin{tabular}{cccccc} \\
        i&j&{\bf T}&$|{\bf R}|$ (\AA)&$N_j$&$J_{5j}^{\bf T}$ (mRy)\\ \hline
        5&1&(0,0,0)&3.856&4&1.155 \\
        5&3&(0,0,0)&3.834&2&0.118 \\
        5&4&(0,0,-1)&3.834&2&0.118 \\
        5&2&(0,0,0)&5.390&2&0.128 \\
        5&4&(0,0,0)&8.533&4&0.045 \\
        5&5&(1,0,0)&5.453&4&0.041 \\
        5&5&(1,1,0)&8.811&4&0.248 \\
        5&5&(0,0,1)&10.780&2&-0.010 \\
        1&3&(0,0,0)&3.834&2&-0.482 \\
        1&4&(0,0,-1)&3.834&2&-0.482 \\
        2&3&(0,0,0)&3.834&2&-0.820 \\
        2&4&(0,0,0)&3.834&2&-0.820 \\
      \end{tabular}
    \end{ruledtabular}
\end{table}

We can see that all of the exchange interactions between the Mn$_{\rm Ge}$ antisite with regular Mn are ferromagnetic with J$_{51}^{\bf 0}$ being the strongest interaction. It is in fact stronger by an order of magnitude than the other ones even though it is not the nearest neighbor. The $J_{52}^{\bf 0}=J_{54}^{(0,0,-1)}$ corresponds to a slightly shorter distance because of the $c/a<2$ but is between opposite spins in the reference structure. 
The exchange interaction between two Mn$_{\rm Ge}$, $J_{55}$ are also ferromagnetic and strongest for the (1,1,0)  direction. Meanwhile, the exchange interactions between the regular Mn atoms on Mn sites stay antiferromagnetic between near neighbor layers but are weaker for the Mn in the plane of the Mn$_{\rm Ge}$.

The reference spin configuration studied here has a net surplus of spin up over spin down Mn and is thus ferrimagnetic. However, the presence of ferromagnetic coupling to the neighboring Mn even the ones of opposite spin in the reference raises the question whether  this  reference system is indeed the  ground state.
It is clear that all the Mn in the MnP plane have ferromagnetic coupling.
As for the net coupling between the adjacent layers, like Mn$_3$, these atoms have ferromagnetic coupling with two Mn$_5$ but still stronger antiferromagnetic coupling to the two Mn$_1$ and even stronger antiferromagnetic coupling with the atoms in the next Mn$_2$ spin-up layer. It thus appears that the overall antiferromagnetic ordering  of alternating spins along the [001] direction is compatible with the exchange interactions found and would be maintained. To test this, a ferromagnetic alignment of all Mn in this cell was calculated and found indeed to have higher energy in the GGA+$U$ model than the AFM configuration. 

\begin{figure*}
	\includegraphics[width=15cm]{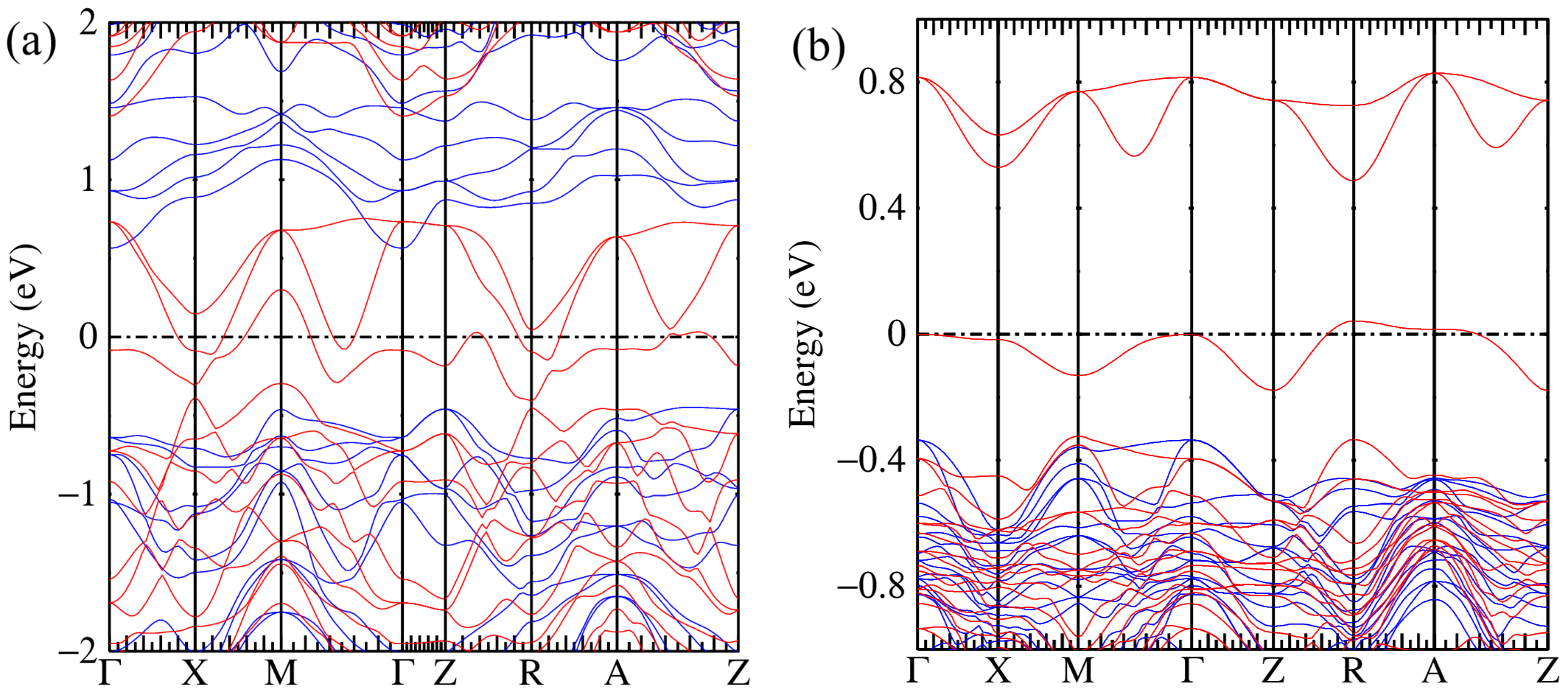}
	\caption{Band structures of the (a) 16 atom cell and (b) 64 atom cell obtained in the QS$GW$ approach. The majority spin bands are shown in red and the minority spin bands are shown in blue. The Fermi energies are shifted to zero. 
		\label{fig-16_64_atom_bands}}
\end{figure*}

Adding this MnP layer, the system also becomes 
metallic. Indeed, the band structure for this 16 atom cell is shown in  Fig. \ref{fig-16_64_atom_bands} (a) and shows the Fermi level crossing through the majority spin bands. Qualitatively this agrees with the observation of metallic behavior in samples of MnGeP$_2$ in which direct evidence was obtained for MnP clusters and for ferromagnetic resonance \cite{Medvedkin2024b}. Although our model does not faithfully represent actual MnP structure precipitates but rather a single layer of tetrahedrally bonded MnP inserted in the otherwise AFM structure, it shows that inserting MnP can make the system ferrimagnetic and metallic. The Curie critical temperature in our model is higher than the observed ones and we expect that quantum corrections might still further increase it. Actual MnP has a $T_C=291$ K close to the observed ferromagnetic temperature in MnGeP$_2$ samples which were shown to contain MnP precipitates \cite{Bardeleben2024}.
From the above analysis, it is clear  that merely inserting a layer of MnP as in our model is not sufficient to make the overall system ferromagnetic. The occurrence of a metallic band structure might induce longer range carrier mediated exchange couplings, but in principle these are already included in our model as the exchange interactions are obtained from the electronic structure. This suggests that to obtain ferromagnetic behavior, larger cluster of actual MnP structure may be required to form as a true secondary phase. 

In the above model we broke the overall stoichiometry by adding net Mn. We also studied a 16 atom cell in which the Mn$_{\rm Ge}$ is compensated by a nearby Ge$_{\rm Mn}$, in other words we create a exchange defect by swapping a Mn and a Ge instead of adding Mn.
In that case, we find that the $J_{53}=-1.393$ mRy of the Mn$_{\rm Ge}$ with its nearest neighbor layers is antiferromagnetic. The exchange interactions between adjacent layer Mn, such as $J_{23}=J_{24}=-1.087$ mRy also stay antiferromagnetic
while the $J_{25}=-0.075$ mRy becomes weakly antiferromagnetic. This again suggests that the AFM ordering of adjacent (001) planes is maintained as in the perfect crystal. There is in this case also a weak $J_{55}^{(110)}=-0.138$ mRy antiferromagnetic coupling and also $J_{55}^{(100)}=-0.060$ mRy coupling, suggesting that in the basal plane antiferromagnetic ordering of the Mn$_{\rm Ge}$ could be favored in a checkerboard pattern doubling the unit cell in two directions. But we should recall that this model of just swapping a Mn and Ge in the basal plane itself is somewhat artificial in the first place.

Next we consider whether more isolated Mn$_{\rm Ge}$ may already introduce ferromagnetic couplings and metallicity.
In this model, we start from the conventional AFM cell, double it in the $x$ and $y$ direction and then replace one Ge by a Mn. The converged magnetic moments keep their spin direction and are about $\pm4.6$ $\mu_B$ on the regular Mn and 3.4 $\mu_B$ on the antisite. The strongest interaction of the Mn$_{\rm Ge}$ is again with its nearest neighbor regular Mn of the same spin and in the same basal plane at a distance of 3.856 \AA, and is ferromagnetic with a value of 1.309 mRy.
The other interactions with near neighbors of opposite spin in the reference structure in the layers just above and below are 0.221 mRy. The interactions with like spin atoms at a distance of 7.667 \AA\ are 0.103 mRy.
These are somewhat larger than in the 16 atom cell but still of similar magnitude. The Mn$_{\rm Ge}$-Mn$_{\rm Ge}$ are farther away and their interactions become negligible, indicating that we approach the isolated defect limit. Since we have  net more spin up than spin down Mn the overall system in the reference state calculated here is strictly speaking ferrimagnetic. In this model, the mean-field estimate of the critical temperature is 419.1 K and the RPA estimate is 256.1 K.  The band structure shown in Fig. \ref{fig-16_64_atom_bands} (b) shows a partially filled defect band in the gap
which also indicates possibly metallic behavior by hole carriers.  As in  the 16 atom cell, the nearest neighbor Mn around each Mn$_{\rm Ge}$ have ferromagnetic interactions, which indicates that a cluster of parallel spins may form. 
However, the exchange interactions between near neighbor regular Mn site atoms in adjacent basal planes stays antiferromagnetic and range from -0.67 to -0.996 mRy. This indicates that the host MnGeP$_2$ maintains an antiferromagnetic ordering in which the small clusters of parallel spin around each Mn$_{\rm Ge}$ are embedded. Assuming the concentration  of Mn$_{\rm Ge}$ antisites to be above the percolation threshold, such a system may overall resemble a ferromagnet but would be better described as a superparamagnet with local clusters of parallel spin, and would presumably have  only small hysteresis, similar to the situation in most dilute  magnetic semiconductors. 

In summary of this section the presence of Mn$_{\rm Ge}$ antisites is found to lead to ferromagnetic coupling between these antisite Mn and the near neighbor regular Mn atoms in the system. This conclusion agrees with the findings of Mahadevan and Zunger\cite{Mahadevan2003} but it only suggest the possible presence of locally  ferromagnetically aligned clusters around excess Mn while exchange defects or Mn-Ge disorder preserves antiferromagnetic coupling. 

Under certain growth conditions, the occurrence of MnP as a secondary phase was established by X-ray diffraction, \cite{Bardeleben2024} indicating that these precipitates were large enough and these samples showed ferromagnetic resonance with a Curie temperature close to that of the known $T_C$ of bulk MnP.  This does not preclude that other mechanisms for ferri or ferromagnetism could occur under different crystal growth conditions closer to ideal MnGeP$_2$. On the other hand it may be difficult to avoid low energy exchange defect or slight deviations from stoichiometry altogether as these defects have a fairly low energy of formation. But while adding Mn$_{\rm Ge}$  produces some ferromagnetic couplings near them,  they are also insufficient to change the overall system into a ferromagnet.  Our calculations suggest the formation of parallel spin clusters near Mn$_{\rm Ge}$ antisites embedded in an  overall antiferromagnetic host.

\section{Conclusions}

Our results in this paper fall in two distinct categories. First, we studied the electronic structure and the derived interband optical transition and magnetic properties of pure chalcopyrite MnGeP$_2$ using the QS$GW$ approach. In the second part, we consider possible defect origins of the experimentally observed ferromagnetism.

In agreement with previous DFT work, we found an antiferromagnetic ground state for pure MnGeP$_2$.  We determined the band structure and optical dielectric function using the QS$GW^{BSE}$ variant of the method in which electron-hole effects are included in the screening of $W$. We find an indirect band gap of about 1.87 eV and  lowest direct gap of 2.44 eV and significant excitonic effects.  
We also found that in a ferromagnetic configuration the gap would be  significantly smaller.
Interestingly, the antiferromagnetic structure is actually an altermagnet, showing spin splittings of the bands along certain {\bf k}-lines. This is because the two magnetic sites of opposite spin in the unit cell are related by a rotation operation. Spin degeneracy is maintained along some symmetry lines in agreement with the group theoretical predictions described in \cite{Smejkal2022}

  We presented a detailed study of the exchange interactions extracted from the
  transverse spin susceptibility calculated in linear response from the QS$GW$
  wavefunctions and eigenvalues, which is then averaged  over atomic sites and mapped onto a Heisenberg-type spin-Hamiltonian following the approach of Kotani and van Schilfgaarde \cite{Kotani2008}. This confirms the dominance of a strong antiferromagnetic coupling of order $-1$ mRy between the two Mn in the primitive cell. The N\'eel temperature is calculated in both mean field and RPA approximations and estimated to be about 173 K within a classical treatment of the spin dynamics. It could  be higher of order 242 K if we include a quantum correction
  factor. We also calculated the spin wave dispersion and the associated dynamical spin susceptibility and find, as expected, linear behavior of the spin waves near the $\Gamma$-point and a maximum spin wave energy of about 58 meV near the Brillouin zone boundary, of the same order of magnitude as in MnO.   
  
  In the second part of the paper, we studied several possible mechanisms which could explain the occurrence of ferromagnetism in real samples in spite of the antiferromagnetic ground state of the perfect crystal. We found that carrier mediated ferromagnetic coupling cannot explain the occurrence of ferromagnetism. While adding electrons or holes in a rigid band structure lowers the main AFM coupling, this does not lead to ferromagnetism because the sign of the coupling does not change.  Even with unrealistically high doping concentrations, the effect is too small to  flip the sign of the coupling. On the other hand, M$_{\rm Ge}$ antisites were found to induce ferromagnetic coupling with nearby regular Mn sites. These interactions were the strongest with the nearest neighbor same spin Mn in the same basal plane and of order 1 mRy, while they are an order of magnitude smaller with the opposite spin nearest neighbors.
  This indicates that clusters of parallel spin are likely to occur near such antisites and could lead to dilute semiconductor type superparamagnetism if these defects occur in a sufficient concentration to be above the percolation limit.
 However, these clusters of locally parallel spins would still be embedded in an overall antiferromagnetic host. 
 On the other hand a MnP  layer inserted inside AFM MnGeP$_2$ was also found to give strong ferromagnetic coupling and with a ferrimagnetic critical temperature larger than that of actual MnP. However, even in this case we argued that the remaining antiferromagnetic couplings between regular Mn sites would  maintain the antiferromagnetic ordering among alternate (001) planes. We also find that these models with Mn$_{\rm Ge}$ antisites have metallic band structure or show partially filled defect bands which will tend to give p-type conduction.  
 While thus far all experimentally realized samples of MnGeP$_2$ show ferromagnetic behavior which has usually been ascribed to a secondary MnP phase, and show metallic behavior, it would be of great interest to realize more perfect defect free MnGeP$_2$ with a semiconducting band structure to test the predicted antiferromagnetism and in fact altermagnetism as well as the spin wave spectra.

 \acknowledgments{This work was supported by the U.S. Department of Energy Basic Energy Sciences (DOE-BES) under Grant No. DE-SC0008933. Calculations made use of the High Performance Computing Resource in the Core Facility for Advanced Research Computing at Case Western Reserve University. J.J. acknowledges support under the CCP9 project Computational Electronic Structure of Condensed Matter [part of the Computational Science Centre for Research Communities (CoSeC)].}
 \bigskip
 
 \centerline{\bf DATA  AVAILABILITY}
 The data that support the findings of this article are publicly available
\cite{datavail}.

\bibliography{Bib/lmto,Bib/dft,Bib/gw,Bib/BSE,Bib/spinchi,Bib/mngep2}
\end{document}